\documentclass[a4paper,11pt]{article}

\usepackage{amsmath}
\usepackage{graphicx}
\usepackage{amsfonts}
\usepackage{amssymb}
\usepackage{diagrams}

\usepackage{color}
\usepackage{hyperref}
\hypersetup{colorlinks}

\definecolor{darkred}{rgb}{1,0,0}
\definecolor{darkgreen}{rgb}{0,0.5,0}
\definecolor{darkblue}{rgb}{0,0,1}
\definecolor{orange}{rgb}{1,0.5,0}
\definecolor{green}{rgb}{0,1,0}
\definecolor{purple}{rgb}{.5,0,1}

\hypersetup{ colorlinks,
linkcolor=darkred,
filecolor=darkgreen,
urlcolor=darkblue,
citecolor=darkblue,
linktocpage=true }

\numberwithin{equation}{section}
\newtheorem{definition}{Definition}[section]

\newtheorem{proposition}[definition]{Proposition}

\newtheorem{remark}[definition]{Remark}

\newcommand{\ie}{{\it i.e.}\ }
\newcommand{\be}{\begin{equation}}
\newcommand{\ee}{\end{equation}}
\newcommand{\bea}{\begin{eqnarray}}
\newcommand{\eea}{\end{eqnarray}}
\newcommand{\bc}{\begin{center}}
\newcommand{\ec}{\end{center}}
\def\ba#1{\begin{array}{#1}\displaystyle}
\newcommand{\ea}{\end{array}}

\newcommand{\beq}{\begin{equation}}
\newcommand{\eeq}{\end{equation}}
\newcommand{\beqa}{\begin{eqnarray}}
\newcommand{\eeqa}{\end{eqnarray}}

\newcommand{\n}{\nonumber\\}
\newcommand{\bi}{\begin{itemize}}
\newcommand{\ei}{\end{itemize}}
\def\mato{\left(\begin{matrix}} 
\def\matf{\end{matrix}\right)}

\def\lt#1{\left#1}
\def\rt#1{\right#1}
\def\t#1{\tilde{#1}}

\def\frc#1#2{\frac{#1}{#2}}

\newcommand{\p}{\partial}

\newcommand{\Pexp}{{\cal P}\exp}

\newcommand{\NN}{{\mathbb{N}}}
\newcommand{\RR}{{\mathbb{R}}}
\newcommand{\CC}{{\mathbb{C}}}
\newcommand{\PP}{{\mathbb{P}}}
\newcommand{\1}{\mbox{\hspace{.0em}1\hspace{-.24em}I}}

\newcommand{\si}{\sigma}
\newcommand{\ep}{\epsilon}

\newcommand{\Tr}{{\rm Tr}}

\newcommand{\halmos}{\rule{1ex}{1.4ex}}
\newcommand{\eproof}{\hspace*{\fill}\mbox{$\halmos$}}

\newcommand{\cB}{{\cal B}}

\newcommand{\cF}{{\cal F}}
\newcommand{\cQ}{{\cal Q}}

\begin{document}

\renewcommand{\thefootnote}{\arabic{footnote}}
\setcounter{footnote}{0}
\newpage
\setcounter{page}{0}

\begin{titlepage}
\begin{center}

{\Large \textbf{The Quench Map in an Integrable Classical Field Theory:\\
Nonlinear Schr\"odinger Equation}}

\vspace{2cm}

{\Large \textbf{Vincent Caudrelier$^{a}$ and Benjamin Doyon$^{b}$} }

\vspace{0.5cm}

\noindent
$^{a}${\footnotesize School of Mathematics, University of Leeds,\\ LS2 9JT Leeds, United Kingdom}
\\[3mm]
\noindent
$^{b}${\footnotesize Department of Mathematics, King's College London,\\ Strand, WC2R 2LS, London, United Kingdom}

\vfill

\begin{abstract}
We study the non-equilibrium dynamics obtained by an abrupt change (a {\em quench}) in the parameters of an integrable classical field theory, 
the nonlinear Schr\"odinger equation. We first consider explicit one-soliton examples, which we fully describe by solving the 
direct part of the inverse scattering problem. We then develop some aspects of the general theory 
using elements of the inverse scattering method. For this purpose, 
we introduce the {\em quench map} which acts on the space of scattering data and represents the change of parameter with fixed field configuration (initial condition).
We describe some of its analytic properties by implementing a higher level version of the inverse scattering method, and we discuss the applications of 
Darboux-B\"acklund transformations, Gelfand-Levitan-Marchenko equations and the Rosales series solution to a related, dual quench problem. 
Finally, we comment on the interplay between quantum and classical tools around the theme of quenches and on the usefulness of the quantization of 
our classical approach to the quantum quench problem. 
\end{abstract}
\end{center}

\vfill

\noindent {\footnotesize {\tt E-mail: v.caudrelier@leeds.ac.uk,\  benjamin.doyon@kcl.ac.uk }}\\


\end{titlepage}

\section{Introduction}

The study of non-equilibrium systems has received a large amount of attention recently, especially in two areas: that of classical stochastic processes and that of quantum many-body problems.
Insisting on Hamiltonian, non-stochastic dynamics, a particular protocol of high interest, which has the potential for experimental observations, 
is that of ``quenches'': abrupt changes in the parameters of the system. This has been studied intensively within the context of extended quantum systems, and, 
especially within the framework of integrability, it led to a large body of theoretical results \cite{qu1}-\cite{Dtherm}, see also the reviews \cite{PSS11} and \cite{EFG15}. These protocols pose many questions, including the properties of the dynamics following a quench, as well as the thermalization and 
(especially in integrable systems) generalized thermalization \cite{GGE1,rigol2,IlieGGE,Dtherm} processes describing the large-time steady state.

It is natural to ask similar questions for Hamiltonian dynamics of extended {\em classical} systems: how does a steady (or perhaps simply evolving) solution to a classical field equation 
(a partial differential equation in space-time) evolve after a parameter of the equation is abruptly modified? 
In the realm of classical integrable PDEs, of particular interest is the nonlinear Schr\"odinger equation (NLSE). It is a prototype of such 
integrable PDEs, and hence has a natural pedagogical status to study new ideas. It also appears 
in the context of nonlinear optics \cite{Ag} and water wave phenomena (although, arguably, it might be more realistic to be able to control the parameters 
to create a quench in a nonlinear optics experiment). In this paper, we will concentrate entirely on the NLSE, and the quench will be with respect to the unique parameter of the NLSE.

Of course, contrary to the situation for many-body quantum systems, the numerical simulation of solutions to classical field equations is much easier, and therefore classical quench dynamics are in principle more accessible numerically. 
In addition, although not posed in this manner, the question of such classical quenches in integrable field theory has been studied for a long time: it 
is ``simply'' a special initial value problem. That is, borrowing the terminology from the quantum arena, an initial state is prepared (the initial condition) that might be 
a special state for a given value of the parameter in the PDE (e.g. a soliton), and is then evolved according to the same PDE but with a different 
value of that parameter. The long-time behaviour of the solution of such initial-value problems is one of the key questions, and
in the context of integrable PDEs, powerful methods exist that are related to the 
inverse scattering method \cite{GGKM,ZS,AKNS}.

Despite this, a general theory of classical quenches, 
within the context of the inverse scattering method, is largely missing.
Indeed, although a quench problem is an initial-value problem, solved in principle a long time ago with the advent 
of the inverse scattering method, one cannot but notice that the direct part of the method has been largely overlooked. The direct part is the mapping from the field to the scattering data (usually understood as a generalization of the Fourier transform), and is essential in solving the initial value problem. 
But it is the inverse part (which actually 
gave its name to the method), corresponding to the inverse mapping, that has mainly promoted the successes of the method:
 {\em given} some assumptions on the form and/or analytic properties of the scattering data, one derives multisoliton solutions and the long-time behaviour of 
 more general solutions
 via the 
nonlinear steepest descent method \cite{DZ}. Except for some deep general results \cite{BC} that cannot easily serve our purposes, 
to our knowledge, the full extent of the bibliography providing explicit results for the direct part is covered by our references \cite{SY}-\cite{GS}. 

It is worth noting that papers \cite{GBC,GS} are extremely recent and precisely motivated by the question of quenches. Therefore, we would like to argue that the question of quenches in 
integrable classical theories could serve as a starting point to stimulate (again) studies in the direct part of the inverse scattering method. In addition, we believe that many ideas recently developed in the quantum context have a clear counterpart in the general theory of classical quenches, and that this interplay may be very fruitful. Indeed, it is well-known that 
integrable systems share the special trait that many techniques and results ``survive'' quantization exactly: in a nutshell, the inverse scattering method becomes the 
quantum inverse scattering method.

From the 
quantum point of view, the fundamental quantities that can be used to understand the post-quench dynamics are the overlaps between eigenstates in the pre-quench and post-quench bases; in \cite{CE13} this idea is developed into the quench action formalism. We will show that the classical 
analog of these overlaps can likewise be subject to a general theory. To achieve this, we propose for this classical analogue the notion of {\em quench map}, that captures the essence of the quench problem in classical integrable theories from the point of 
view of the inverse scattering method.
The map acts on the space of scattering data and represents a change of parameters (thus a change of the scattering transform) with fixed field configuration (\ie 
a prescribed initial condition). 
The quench map is the basic object for describing classical quenches. We study it through three explicit examples and then in more generality 
by developing a scattering problem that one could interpret as a ``higher-level'' version of the usual Lax pair approach. This allows us to 
establish certain analytic properties that the quench map should satisfy.

Finally, we study a ``dual quench'' protocol, where the field is abruptly changed while the scattering data are fixed. This is related to the quench map by 
intertwining with the scattering transform, and has an interpretation, by analogy with the quantum case, as the effect of a slow adiabatic change of the parameter 
of the NLSE (hence is, in a sense, the dual protocol to that of quenches). We provide some observations about the role of the so-called B\"acklund transformations, 
the Rosales series representation and the Gelfand-Levitan-Marchenko equation in relation to the dual quench. These considerations are perfectly natural from the 
classical point of view but perhaps less so from the quantum point of view. Nevertheless, we argue that our results may relate to ideas \cite{SFM} involving the 
Zamolodchikov-Faddeev algebra.

%
%

The paper is organized as follows. In Section \ref{Lax}, we briefly recall the usual setup for the inverse scattering method for the nonlinear Schr\"odinger 
equation. With this at hand, we then introduce in Section \ref{sectqm} the main object of study for a classical quench: the quench map. This Section 
also contains some very general properties of this map and its natural connection with the (direct part of) the inverse scattering method. Section \ref{direct_part}
contains three fully worked out examples of computation of the form of the quench map at specific points in scattering data space. Although some common features 
appear in those three particular examples (like a special link to Gauss hypergeometric function), the involved tools do not appear 
to be easily generalizable for a more general discussion. Therefore, in Section \ref{sectgen}, we investigate the general problem by going back to the fundamentals 
and implementing a ``higher-level'' version of the inverse scattering method, whereby the quench scattering data can be characterised via the analytic properties 
of wavefunctions of a ``level-one'' auxiliary problem. The latter involves the original Lax matrix $U$ of the ``level-zero'' auxiliary problem dressed by the 
Jost solutions of that level-zero problem. In Section \ref{real_quench}, we transpose the quench problem into ``real space'': we consider the dual quench problem, 
where the field changes while the scattering data are fixed. This has a natural connection with the quench map problem via a simple 
intertwining relation. Therefore this allows us to 
connect the quench map problem to several well-known tools of the inverse scattering method: B\"acklund transformations, Gelfand-Levitan-Marchenko equations and 
the related Rosales expansion. All these have natural quantum counterparts and we comment on the connection between our classical approach and previous studies 
in the literature on the quantum quench problem. The last Section contains our conclusions and outlook. 

\section{Lax pair representation and inverse scattering method for the NLSE}\label{Lax}

In this Section we recall the basic theory of the NLSE; this serves to introduce our notation.
We consider the NLSE for the complex-valued function $q$ of the two independent variables $x$ and $t$, where the strength of the cubic nonlinearity is parametrized as $c^2$:
\be
\label{eq:NLS}
i\partial_t q+\partial_x^2 q-2c^2|q|^2q=0\,.
\ee 
The case $c\in\RR^+$ is the defocusing case while the case $c\in i\RR^+$ is the focusing case; $c=0$  is simply the noninteracting or free case. The equation appears as 
the compatibility condition of an overdetermined linear system characterized by the 
so-called auxiliary problem for $\Psi(x,t)\in\CC^2$ \cite{ZS}
\bea
\begin{cases}
\label{pb_aux1}
\Psi_x=(-ik\sigma_3+W(x,t))\Psi\,,\\
\label{pb_aux2}
\Psi_t=(-2ik^2\sigma_3+P(x,t,k))\Psi\,,
\end{cases}
\eea
where the indices indicate partial derivatives, and the $2\times 2$ matrix-valued functions $W(x,t)$ and $P(x,t,k)$ form the Lax pair \cite{Lax} and are given by
\bea
\label{lax_pair}
W(x,t)=c\left(\begin{array}{cc}
0 & q(x,t)\\
\,q^*(x,t) & 0
\end{array}\right)~~,~~P(x,t,k)=2k\,W-iW_x\,\sigma_3-iW^2\,\sigma_3\,.
\eea 

In order to discuss the quench problem  we introduce the scattering data associated to the auxiliary problem \eqref{pb_aux1}.
The scattering data can be encoded into a matrix-valued function $S(k)$  (the scattering matrix), as a function of a spectral parameter $k\in\RR$, whose 
extension to $k\in\CC$ has particular analytic properties. It forms the basis of the inverse scattering method (ISM) \cite{GGKM,ZS,AKNS}. In particular, the map from a field configuration $x\mapsto q(x,0)$ at a given time (here $t=0$), to the scattering data $k\mapsto S(k)$ -- the direct scattering transform -- 
can be seen as a generalization of the Fourier transform that is well adapted to the nonlinear integrable partial differential equations. Indeed, under 
time evolution, the scattering data evolve in a simple manner. This simple evolution in principle provides the solution to the problem, as the inverse map -- 
the inverse scattering transform -- can be obtained by two main techniques, the Gelfand-Levitan-Marchenko equations or the Riemann-Hilbert factorisation problem 
(see e.g. \cite{FT}).

For the NLSE, the details of the scattering transform depend on the class of functions $q(x,0)$ (and more generally $q(x,t)$) in which we wish to consider the problem. The two classes we consider in this paper correspond to the 
Schwarz class, and the so-called finite density class \cite{FT}. In the former class, the rapidly decreasing case, we impose $q$ as well as all of its higher $x$-derivatives to vanish as $|x|\to\infty$ faster than any polynomial. In the latter class, we impose instead
 \be
 \label{asymtotic}
 \lim_{x\to\pm\infty}q(x,t)=\rho e^{i\theta_\pm(t)}~~,~~\rho>0\,,
 \ee
in the sense that $q-\rho e^{i\theta_\pm}$ is of Schwarz type as $x\to\pm\infty$. Note that this is compatible with \eqref{eq:NLS} provided 
$\theta_\pm(t)$ is linear in $t$ and the difference $\theta_+-\theta_-$ is constant. Therefore, in this case, it is convenient to 
take $\theta_-=0$ and $\theta_+=\theta$ a constant in $[0,2\pi)$, and to consider the gauged transformed NLSE
\be
\label{NLS_finiteD}
i\partial_t q+\partial_x^2 q-2c^2(|q|^2-\rho^2)q=0\,,
\ee 
related to the NLSE via $q(x,t)\mapsto e^{-2ic^2\rho^2t}q(x,t)$. 
This is incorporated in the auxiliary problem by simply changing $P$ to $P+ic^2\rho^2\sigma_3$ while keeping everything 
else unchanged. In particular, this does not affect the form of the $x$-part of the auxiliary problem, which is the important 
part for computing the scattering data from a given initial condition. The special case $\rho=0$ of the finite density class recovers the Schwarz class.

The scattering data is then computed by considering two Jost solutions $\Psi^\pm$ of the $x$-part of the auxiliary problem at fixed time
with appropriate asymptotics as $x\to \pm \infty$. 
Let us denote
\be
W_\pm:=\lim_{x\to\pm\infty}W(x,0)\,,
\ee
and 
\be
E_\pm(x,k):=P_\pm e^{-i\Lambda_\pm(k)x}
\ee
the solution of
\be
E_x+(ik\sigma_3-W_\pm)E=0\,,~~E(0,k)=\1_2\,,
\ee
where we have assumed that $ik\sigma_3-W_\pm$ is diagonalizable to $i\Lambda_\pm(k)$ via the matrix $P_\pm$ and where $\1_2$ is the unit $2\times 2$ matrix. Then,
the Jost solutions $\Psi^\pm=\Psi^\pm(x,k)$ are defined as the unique solutions to
\be
\label{sol_Psi1}
\Psi_x+ik\sigma_3\Psi=W(x,0)\Psi
\ee
satisfying the asymptotic conditions
\be
\label{sol_Psi2}
\lim_{x\to\pm\infty}E^{-1}_\pm(x,k)\Psi^\pm(x,k)=\1_2\,.
\ee
The scattering matrix $S(k)$ is then the matrix connecting the two Jost solutions according to 
\be
\Psi^+(x,k)=\Psi^-(x,k)\,S(k)\,.
\ee
In practice, we can compute it as follows
\be
\label{compute_S}
S(k)=\lim_{x\to-\infty}E^{-1}_-(x,k)\Psi^+(x,k)\,.
\ee
The spectral parameter takes values in $k\in\RR_\rho$, where $\RR_\rho$ is $\RR$ in the rapidly decreasing case, and 
is $\RR$ minus a fixed interval depending on $\rho$ in the finite density case \cite{FT}. 
The task is further simplified by noting that there is a symmetry of the auxiliary problem that follows from the property
\be
W^*=\sigma\,W\,\sigma\,,
\ee
where
\be
	\sigma = \left(\begin{matrix}
	0 & c\,/\,|c| \\ c^*\,/\,|c| & 0 \end{matrix}\rt)\,.
\ee
This implies the following form for $S(k)$:
\be\label{Sab}
S(k)=\left(\begin{array}{cc}
 a^*(k)& b(k)\\
 \frac{c^*}{c}\,b^*(k) & a(k) \end{array}\right)\,.
\ee 
Combined with the fact that $S(k)$ has unit determinant, 
this involution property has a more transparent form in the focusing case,
\be
S^\dagger(k)=S^{-1}(k)\quad (k\in\RR_\rho)\,,
\ee
showing that $S(k)$ is in the group $SU(2)$ for each $k\in\RR_\rho$. The details of the involution on the spectral parameter $k$ depend on the class we are working in and will be clear 
in the examples below. Further details can be found in \cite{FT}. The main observation for our purposes is that it is enough to compute 
$a$ and $b$ in order to completely characterize the direct scattering transform.
Finally, the scattering data $S(k)$ evolves in a simple manner under the time evolution determined by the NLSE \eqref{eq:NLS} (or \eqref{NLS_finiteD} in the finite-density case):
\be\label{evol}
	S(k;t)=e^{2ik^2t\sigma_3}S(k)e^{-2ik^2t\sigma_3}\,.
\ee

\section{General ideas and the quench map} \label{sectqm}

Following the usual implementation of a quench, one first assumes that the field configuration $q(x,t)$ is a stationary, or in some way natural, solution to the 
NLSE, with parameter $c$, for all times $t<0$. In quantum problems, one often takes the ground state of a given many-body Hamiltonian; in the present case, 
one may instead assume the presence of one or a finite number of solitons, or a stationary extrapolating solution in the finite-density case (the zero-field 
solution $q=0$, in the Schwartz case, would give a trivial quench). Then, an abrupt change $c\mapsto c'$ of the parameter is performed. This is implemented 
by considering $q(x,0)$ as the initial condition for a new time evolution, corresponding to the NLSE with this modified parameter. The solution $q(x,t)$ for all $t>0$ is determined by solving the NLSE with parameter $c'$, from the initial condition $q(x,0)$ (which, we emphasize, was assumed to be a natural solution to the NLSE with parameter $c$).

Since the time evolution \eqref{evol} is simple in the space of the scattering data, it is convenient to recast the quench problem to that space. Denote by
\be
	{\cal F}_c:\{x\mapsto q(x)\} \to \{k\mapsto S(k)\}
\ee
the direct scattering map with parameter $c$, from field configurations to scattering data, as defined by \eqref{compute_S}. Since $q(x,t),\,t<0$ is a natural solution to the NLSE with coupling $c$, the scattering data $S(k)=S(k,c):={\cal F}_c[q(\cdot,0)](k)$ is assumed to be known and simple, and the time evolution of $S(k,c)$ for $t<0$ is given by \eqref{evol}. The time evolution of $S(k,c)$ for $t>0$ is however not simple, as it is performed by using the NLSE with parameter $c'\neq c$. In order to solve the quench problem in the space of scattering data, we need to obtain $S(k,c'):={\cal F}_{c'}[q(\cdot,0)](k)$. Once $S(k,c')$ is evaluated, the quench problem is in principle solved, as the evolution of the new scattering data for $t>0$ is again given by \eqref{evol}. The inverse scattering method, which gives the inverse transform ${\cal F}_{c'}^{-1}$, then allows one to recover $q(x,t)$ for $t>0$.

Let us introduce the {\em quench map} as the bijective map\footnote{Bijectivity of $\cF_c$ is a difficult problem in general but is 
known to hold for the classes of functions that we consider in this paper.}
\beq\label{Qccp}
	{\cal Q}_{c',c} = {\cal F}_{c'}\circ {\cal F}_c^{-1}
\eeq
on the space of scattering data. The full time evolution in this space is then
\be
	S(k;t) = \begin{cases}
	e^{2ik^2t\sigma_3}\,S(k)\, e^{-2ik^2t\sigma_3} & t<0\\
	e^{2ik^2t\sigma_3}\,{\cal Q}_{c',c} [S](k)\,
	e^{-2ik^2t\sigma_3} & t>0
	\end{cases}
\ee
where $S(k) = S(k,c)$. The quench problem at the time $t=0$ can be visualised as the commutative diagram in Figure \ref{quench_scattering}.
\begin{figure}[htp]
\def\Id{{\rm id}}
\begin{diagram}
q(x,0) & & \rTo^{\cF_{c}} & & S(k,c)  \\
\dTo^{\Id} & &  & & \dTo_{{\cal Q}_{c',c}} \\
q(x,0) & &  \rTo^{\cF_{c'}} & & S(k,c') \\
\end{diagram}
\caption{The quench problem in scattering data space.}
\label{quench_scattering}
\end{figure}
Hence, the knowledge of the quench map is in principle enough to solve the quench problem.
In Section \ref{sectgen}, we explore the relation between the quench map and the inverse scattering method, and show that it leads to a sort of \textit{higher-level auxiliary problem} that captures some of the properties of 
the quench map ${\cal Q}_{c',c}$.

In the quantum quench problem, the important objects are the overlaps between the eigenstates of the initial and final Hamiltonians. It is natural to interpret 
the quench map $\cQ_{c',c}$ as the classical analogue of such overlaps. Indeed, the field $q(x)$ is interpreted as the quantum state, and the scattering data 
$S(k,c)$ ($S(k,c')$) as the coefficients of the state in the basis of eigenstates of the initial (final) Hamiltonian. The overlaps between the eigenstates of the 
initial and final Hamiltonians allow one to go from one basis to the other, and thus their classical analogue is the quench map.

\begin{remark}
The quench map has obvious group properties 
\be
{\cal Q}_{c,c}={\rm id}~~,~~{\cal Q}_{c'',c'}\circ{\cal Q}_{c',c}={\cal Q}_{c'',c}~~,~~{\cal Q}^{-1}_{c',c}={\cal Q}_{c,c'}.
\ee
For any given $c_0$, we may define ${\cal Q}_c:={\cal Q}_{c,c_0}$ and we have 
${\cal Q}_{c',c} = {\cal Q}_{c'}\circ {\cal Q}_{c}^{-1}$. In the case of the NLSE, we may always rescale the field 
$q\to \sqrt{|c|}q$ when $c\neq 0$, so that we can always consider the quench problem with $|c_0|=1$: the {\it normalized} quench problem.
\end{remark}

From the point of view of the inverse scattering method, it is important to realise that the quench problem is completely solved if we can implement explicitely 
the \textit{direct part} of the inverse scattering method, that is, if we can compute the scattering data $S(k,c)$ for fixed but arbitrary $c$, given 
an initial condition $q(x,0)$. Indeed, this amounts to finding $\cF_c$ for all values of $c$ and then the map $\cQ_c$ for the normalized quench problem is obtained from
\be
	{\cal Q}_{c} = {\cal F}_{c}\circ {\cal F}_{c_0}^{-1}\,.
\ee

Historically, the direct part has been less studied than the inverse part of 
the method.
However, for certain classes of initial conditions, 
the direct part can be solved in closed form. To our knowledge, the following references 
exhaust all such studies in the case of NLSE. 
The first occurence is \cite{SY} and, independently it seems \cite{Miles}, for the focusing case with vanishing boundary conditions and 
for a choice of initial data corresponding to a one-soliton solution. Recently, other forms of the initial data were considered 
in \cite{TV} also in the focusing case with vanishing boundary conditions. Much more recently, the same question was solved in 
\cite{GBC} in the defocusing, finite density case and for a choice of initial data corresponding to a soliton solution.
While finalizing this paper, we discovered  
\cite{GS} that tackles the quench problem for a variety of integrable classical field theories, but restricted to very specific initial conditions. 
In the following 
Section, we reproduce the original approach of \cite{SY,Miles} to obtain $\cF_c$ and adapt it to obtain the results presented in \cite{GBC} (now 
also available in the very recent preprint \cite{GS}). We also give a new example corresponding to the focusing case at finite density. All three cases cases share common mathematical features but their 
physical relevance is quite different.

\section{Calculations of the quenched scattering data in the NLSE}\label{direct_part}

In this Section we set out to compute the scattering data $S(k,c)$ explicitely as in \eqref{compute_S} in three different situations: 
rapidly decreasing focusing NLSE, finite density focusing and defocusing NLSE. In each case the initial condition is chosen to be precisely of soliton type
for the normalized value of the paramter ($c_0=i$ or $c_0=1$). 
This gives us the value of the quench map at three specific points in the scattering data space.

\subsection{Rapidly decreasing case}\label{decreasing}

Let us compute the scattering data explicitly starting from the initial data 
\be
\label{1_soliton}
q(x,0)=A\,\frac{e^{i\left(V\,x+\phi_0\right)}}{\cosh\left(A(x-x_0)\right)}\,,
\ee
which corresponds to a one-soliton solution of the focusing NLSE ($c_0=i$) with amplitude $A$, velocity $V$ and some arbitrary phase and position shifts $\phi_0$ and 
$x_0$. The latter can be set to zero without loss of generality. 
The case $V=0$ was considered in \cite{Miles} and restricted there to the focusing case $c\in i\RR$. 
Here, we make no such restrictions since it turns out that the results hold in general. 
For this case, we have $W_\pm=0$ and hence 
\be
\lambda_\pm(k)=k\sigma_3~~,~~P_\pm=\1_2
\ee
so that \eqref{compute_S} yields
\be
\left(\begin{array}{cc}
b(k,c)\\
a(k,c)
\end{array}\right)=\lim_{x\to-\infty}\left(\begin{array}{cc}
\phi^+_{12}(x,k)\\
\phi^+_{22}(x,k)
\end{array}\right)\,,
\ee
where
$\phi_+(x,k)=e^{-ikx\sigma_3}\Psi^+(x,k)$. 
Let us set $z(x)=\frac{1}{2}(1-\tanh Ax)$ and
\bea
\label{def_F}
F(z)=\phi^+_{22}(x,k)\,.
\eea
This yields,
\bea
\label{def_F_suite}
\phi^+_{12}(x,k)=\frac{e^{2ikx}}{cq^*(x,0)}\frac{dz}{dx}F^{'}(z)\,.
\eea
Inserting in $\partial_x\phi^+_{12}(x,k)=cq(x,0)e^{2ikx}\phi^+_{22}(x,k)$, we get, after some algebra involving standard hyperbolic identities
\bea
z(1-z)F^{''}(z)+\left[\frac{1}{2}-\frac{i}{A}\left(k+\frac{V}{2}\right)-z\right]F^{'}(z)-c^2F(z)=0\,.
\eea
Therefore, one finds that $F$ is the solution of the hypergeometric differential equation 
\bea
z(1-z)F^{''}(z)+\left[\gamma-(\alpha+\beta+1)z\right]F^{'}(z)-\alpha\beta F(z)=0\,.
\eea
with 
$\beta=-\alpha$, $\alpha^2=-c^2$, $\gamma=\frac{1}{2}-\frac{i}{A}\left(k+\frac{V}{2}\right)$
. The change $\alpha\to -\alpha$ has no effect so we can set $\alpha=ic$.
The boundary conditions coming from
\be
\lim_{x\to\infty}\phi^+_{22}(x,k)=1~~,~~\lim_{x\to\infty}\phi^+_{12}(x,k)=0\,,
\ee
read
\bea
\label{BC0}
F(0)=1~~,~~\lim_{z\to 0}z^{\frac{1}{2}-\frac{i}{A}(k+\frac{V}{2})}\,F^{'}(z)=0\,.
\eea
Therefore, we find 
\be
F(z)=\mathbin{_2F_1}\left(ic,-ic,\frac{1}{2}-\frac{i}{A}\left(k+\frac{V}{2}\right);z\right)\,.
\ee
Taking the limit of the functions in \eqref{def_F}-\eqref{def_F_suite} as $x\to -\infty$, \ie $z\to 1$, we obtain after simplification,
\bea
a(k,c)&=&\frac{\Gamma\left(\frac{1}{2}-\frac{i}{A}\left(k+\frac{V}{2}\right) \right)^2}
{\Gamma\left(\frac{1}{2}-ic-\frac{i}{A}\left(k+\frac{V}{2}\right) \right)\Gamma\left(\frac{1}{2}+ic-\frac{i}{A}\left(k+\frac{V}{2}\right) \right)}\,,\\
b(k,c)&=&-\frac{\sin(i\pi c)}{ ic }\frac{1}{\cosh\frac{\pi}{A}\left(k+\frac{V}{2} \right)}\,,
\eea
where $\Gamma$ is the Gamma function. When $c=i$, we consistently recover the scattering data corresponding to the one-soliton solution of NLSE, that is
\bea
b(k,i)=0~~,~~a(k,i)&=&\frac{k-k_1}{k-k_1^*}~~,~~k_1=-\frac{V}{2}+\frac{iA}{2}\,.
\eea
Therefore, we get the value of the quench map $\cQ_{c,i}=\cQ_c$ at the ``point'' $(a(k),b(k))=\left(\frac{k-k_1}{k-k_1^*},0\right)$,
\bea
\cQ_c\left(\frac{k-k_1}{k-k_1^*},0\right)=\left(a(k,c),b(k,c)\right)\nonumber\qquad\qquad\qquad\qquad\qquad\qquad\qquad\qquad\qquad\qquad\qquad\\
=\left(\frac{\Gamma\left(1-\frac{i}{A}\left(k-k_1\right) \right)^2}
{\Gamma\left(1-ic-\frac{i}{A}\left(k-k_1\right) \right)\Gamma\left(1+ic-\frac{i}{A}\left(k-k_1\right) \right)},
\frac{\sin(i\pi c)}{ ic }\frac{i}{\sinh\frac{\pi}{A}\left(k-k_1 \right)}\right).
\eea

Thanks to these explicit expressions, we can investigate in detail the effect of a quench. As noted above, although historically \cite{Miles} only considered the focusing 
case, $c\in i\RR$ here, it is clear that $a(k,c)$ and $b(k,c)$ can be analytically continued to other values of $c$. In particular, we can investigate a 
quench that {\it changes the nature of the system} \ie a quench from the focusing to the defocusing regime. To our knowledge, this feature has never been explored before. 

\paragraph{Focusing to focusing quench} In this case, $c\in i\RR$ and we set $c=i\nu$, $\nu>0$. The system is quenched for $\nu\neq 1$.
Of all these values, we see that the case where $\nu=n\in\NN$ is special: $b(k,in)=0$ meaning that $q(x,t)$ is reflectionless \ie of solitonic type. 
The corresponding form 
of $a(k,in)$ is
\be
a(k,in)=\prod_{j=1}^n\frac{k-k_j}{k-k_j^*}~~,~~k_j=-\frac{V}{2}+\frac{(2j-1)iA}{2}\,.
\ee
Note that only the amplitudes change while the velocity remains the same for each zero of $a(k,in)$: the corresponding solution is a bound state of $n$ 
solitons moving together with the same velocity and with a hierarchy of equally spaced amplitudes, separated by $A$. 
In general, for non integral values of $\nu$, the zeros of $a(k,i\nu)$ are located at those values of $k=k_R+ik_I$, $k_I>0$ such that 
the argument of one the Gamma functions in the denominator is a negative integer $-n$, $n=0,1,2,\dots$. This yields, remembering that $k_I>0$, only the 
following possibility
\be
\begin{cases}
k_R=-\frac{V}{2}\,,\\
k_I=-A(n+\frac{1}{2}-\nu)\,.
\end{cases}
\ee
So if $0<\nu<1$, we see that $a(k,i\nu)$ has no zero in the upper-half plane. This means that the initial soliton desintegrates 
into a purely radiative solution controlled by the continuous scattering data $\rho(k,c)=\frac{b(k,c)}{a(k,c)}$, $k\in\RR$. This is reminiscent of 
the well-known 
fact that there is a threshold for the value of the coupling in order to get a soliton in the focusing case.
If the quench is to a value of $c$ such that $N< \nu <N+1$ for some $N\ge 1$, then $a(k,i\nu)$ has $N$ zeros for $n=0,\dots,N-1$, all having the same 
real part. The corresponding solution consists of the abovementioned bound state of solitons together with an additional radiative part that decays 
at large times.

\paragraph{Focusing to defocusing quench}

In this case, we assume that $c>0$. Performing the same analysis to find the possible zeros of $a(k,c)$ in the upper half-plane, 
we find no solutions. 
This is consistent with the known fact that the rapidly decreasing defocusing case does not support discrete eigenvalues leading to 
solitons. We also see that $b(k,c)$ is never zero in that case. So, we find that a quench to a positive
value of the coupling destroys the initial soliton and leaves only a purely radiative solution controlled by $\rho(k,c)=\frac{b(k,c)}{a(k,c)}$, $k\in\RR$. 
We note that there is an interesting expression of the solution in that case, which is a special case of the 
perturbative expansion discovered by Rosales in \cite{Ros}. 
It can also be derived using the Neumann series solution to the Gelfand-Levitan-Marchenko equations for the reconstruction of $q(x,t)$ from the 
scattering data. This is the object of Section \ref{review_Rosales}.

\paragraph{Focusing to free quench}

It is interesting to study the special limit when the coupling goes to zero \ie the limiting case of the free Schr\"odinger equation. Solitons do not 
exist in this regime since they are purely nonlinear objects. Moreover, the inverse scattering method is known to boil down to the
 usual Fourier transform in this case. So we expect $a(k,0)$ to be simply $1$ and $b(k,0)$ to be essentially the Fourier transform of $q(x,0)$.
 In this limit, it is readily checked that $a(k,0)=1$ and 
 \be
 b(k,0)=-\frac{\pi}{\cosh\frac{\pi}{A}\left(k+\frac{V}{2} \right)}\,.
 \ee
 This is precisely the Fourier transform of $q(x,0)$, upon taking into account the change $k\to 2k$ coming from the normalization of our auxiliary problem and 
 an overall sign coming from our definition of the scattering matrix from $\phi^+$ as $x\to\infty$. Explicitly, we can keep track of the small $c$ limit by writing 
 \be
 \Psi^+(x,k)=e^{-ikx\sigma_3}( \1_2+c K(x,k)+O(c^2))\,,
 \ee
 and inserting in \eqref{pb_aux1}. Keeping the first order in $c$, we get
 \be
 K_x(x,k)=\mato 0 & e^{2ikx} q(x,0) \\ e^{-2ikx}  q^*(x,0) & 0 \matf\,.
 \ee
 This yields, using the normalization of $\Psi^+$ as $x\to\infty$,
 \be
 K(x,k)=\mato 0 & -\int_{x}^\infty e^{2ik\xi} q(\xi,0)\,d\xi \\ -\int_{x}^\infty e^{-2ik\xi}  q^*(\xi,0)\, d\xi & 0 \matf\,.
 \ee
The entry $(12)$ in this matrix coincides with $b(k,0)$ as $x\to-\infty$ according to the definition of $S(k,c)$ and is indeed seen to be exactly the 
Fourier transform of the initial data (up to the minor modifications already mentioned).

\subsection{Finite density case with repulsive nonlinearity}\label{finite_rep}

In this case, given the asymptotic behaviour \eqref{asymtotic}, we see that 
\be
W_-=c\mato
0 & \rho \\ 
\rho & 0 
\matf~~,~~W_+=e^{i\frac{\theta}{2}\sigma_3}\,W_-\,e^{-i\frac{\theta}{2}\sigma_3}\,.
\ee
As a consequence, 
\be
\Lambda_+(k)=\Lambda_-(k)=\mu(k)\sigma_3\,,
\ee
where $\mu$ satisfies 
\be
\label{eq_mu}
\mu^2=k^2-c^2\rho^2\,,
\ee
and
\be
P_-=e^{-i\frac{\theta}{2}\sigma_3}\,P_+=\mato
1 & \frac{i(\mu-k)}{c\rho} \\ 
-\frac{i(\mu-k)}{c\rho} & 1 
\matf
\ee
Let us discuss the quench problem using as initial data the following one-soliton solution of \eqref{NLS_finiteD} with $c=1$ \cite{FT}
\be
q_{FD}(x,t)=\rho\frac{1+e^{i\theta}e^{\nu (x-vt)}}{1+e^{\nu (x-vt)}}\,,
\ee
with $\rho>0$ and,
\be
v=-2c\rho\cos\frac{\theta}{2}~~,~~\nu=2c\rho \sin\frac{\theta}{2}\,.
\ee
taken at $t=0$. Hence, the initial data read
\be
\label{initial_finite}
q(x,0)=\rho\frac{1+e^{i\theta}e^{\nu x}}{1+e^{\nu x}}~~,~~\nu>0\,.
\ee

This example was discussed in \cite{GBC} but the derivation of the quenched scattering data was skipped. In the process of completing this paper, 
we discovered the very recent and interesting paper \cite{GS} that provides such details and also analyses other models (KdV and sine-Gordon). Numerical 
results are also provided. 
For self-containedness, and to illustrate our general scheme, we present this case too with our notations. 
Since we work with $\Psi^+$, it is convenient to change to the basis where $W_+$ is diagonal, so we set
\be
\Phi^+(x,k)=P^{-1}e^{-i\frac{\theta}{2}\sigma_3}\Psi^+(x,k)\,.
\ee
The auxiliary problem can again be mapped to the hypergeometric differential equation by using
\be
w(x)=\frac{1}{2}(1-\tanh \frac{\nu x}{2})\,,
\ee
and 
\be
\varphi(w)=\left(\begin{array}{c}
\Phi^+_{12}(x,k)\\
\Phi^+_{22}(x,k)
\end{array}\right)\equiv \left(\begin{array}{c}
\varphi_1(w)\\
\varphi_2(w)
\end{array}\right)\,.
\ee
Noting that 
\be
q(x,0)=\rho(w+e^{i\theta}(1-w))~~,~~\frac{dw}{dx}=-\nu w (1-w)\,,
\ee
we obtain
\be
\frac{d\varphi}{dw}=\frac{i\mu}{\nu}\left(\frac{1}{w}\sigma_3+ \frac{1}{1-w}Q(\theta)^{-1}\sigma_3Q(\theta) \right)\varphi\,,
\ee
where
\be
Q(\theta)=\left(\begin{array}{cc}
\beta^*&-\alpha\\
-\alpha & \beta
\end{array}\right)~~,~~\alpha=\frac{c\rho}{\mu}\sin\frac{\theta}{2}~~,~~\beta=\cos\frac{\theta}{2}+i\frac{k}{\mu}\sin\frac{\theta}{2}\,.
\ee
Setting 
\be
\varphi(w)= \left(\begin{array}{c}
M_1(w)\\
M_2(w)
\end{array}\right)\left(\frac{1-w}{w}\right)^{i\frac{\mu}{\nu}}\,,
\ee
we obtain after some algebra, and using the relation $\beta\beta^*-\alpha^2=1$,
\be
w(1-w)\frac{d^2M_2}{dw^2}-\left(w+\frac{2i\mu}{\nu}\right)\frac{dM_2}{dw}+c^2\,M_2=0\,.
\ee
From our normalisation, $M_2(0)=1$ and we get 
\be
M_2(w)=\mathbin{_2F_1}\left(c,-c,-\frac{2i\mu}{\nu};w\right)\,,
\ee
which in turn yields,
\be
M_1(w)=\frac{\alpha}{\beta^*}\left[\mathbin{_2F_1}\left(c,-c,-\frac{2i\mu}{\nu};w\right)-(1-w)\mathbin{_2F_1}\left(1+c,1-c,1-\frac{2i\mu}{\nu};w\right) \right]
\ee
Therefore, the coefficients $a(k,c)$, $b(k,c)$ are found as
\be
\left(\begin{array}{cc}
b(k,c) \\
a(k,c)
\end{array}\right)=\lim_{w\to 1}
\left(\begin{array}{cc}
\left(\frac{1-w}{w}\right)^{\frac{2i\mu}{\nu}}(\beta^* M_1(w)-\alpha M_2(w))\\
\beta M_2(w)-\alpha M_1(w)
\end{array}\right)\,,
\ee
and read, after simplification,
\bea
a(k,c)&=&-\frac{(\frac{2i \mu}{\nu}-c)(\frac{2i \mu}{\nu}+c)}{\beta^*\left(\frac{2i \mu}{\nu} \right)^2}\frac{\Gamma^2(1-\frac{2i \mu}{\nu})}
{\Gamma(1-\frac{2i \mu}{\nu}-c)\Gamma(1-\frac{2i \mu}{\nu}+c)}\,,\\
b(k,c)&=&\frac{\sin \pi c}{\pi c}\left(\frac{\sinh \frac{2\pi \mu}{\nu}}{\frac{2\pi \mu}{\nu}} \right)^{-1}\,.
\eea
These formulas are deceptively similar to the rapidly decreasing case. But the effect of the nonzero boundary conditions is captured by the important relation 
\eqref{eq_mu}. As before, thanks to these expressions, one can analyse the effect of a quench that does not necessarily preserve the nature of the system.

\paragraph{Defocusing to defocusing quench}

In this context, integer values of $c$ produce reflectionless solutions $b(k,n)=0$. The corresponding coefficient $a(k,n)$ becomes
\be
a(k,n)=\frac{1}{\beta^*}\prod_{m=1}^n\frac{2\mu-im\nu}{2\mu+i(m-1)\nu}\,,
\ee
showing the location of the zeros (in $\mu$). They correspond to discrete values of $k$ in the gap $[-c^2\rho^2,c^2\rho^2]$. 
For noninteger values of $c$, the quench solution combines the previous special multisoliton with a radiative background decaying at large times.
More details can be found in \cite{GS}. 

\paragraph{Defocusing to focusing quench}

If $c=i\gamma$, $\gamma>0$, we immediately see that $b(k,c)$ is never zero so that the original reflectionless soliton desintegrates into a radiative solution.
This is consistent with the fact that $a(k,c)$ cannot have zeros in that case.

\subsection{Finite density case with attractive nonlinearity}

For our final example, we present a quench in a novel situation: the finite density case with {\it attractive} nonlinearity, corresponding to the case $c_0=i$. 
In \cite{BK}, the inverse scattering method and solitons solutions for this case have been studied in detail and a good review of previous results 
is given. We use the following one-soliton initial profile corresponding to the solution rederived by the authors of \cite{BK} 
as a particular case of their general multisoliton formula. Setting $t=0$ as well as $\xi=\phi=0$ in eq $(4.1)$ of \cite{BK}, we consider
\be
q(x,0)=\frac{\cosh(Ax+B)-iA}{\cosh(Ax+B)}\,,
\ee
where
\be
A=Z-1/Z\,,~~\tanh B=\frac{C^2-4Z^2A^2}{C^2+4Z^2A^2}\,,~~C=Z+1/Z\,,~~Z>1\,.
\ee
This initial profile corresponds to a stationary soliton on a nonzero constant background (the density $\rho=1$ here) and is controlled by one parameter $Z>1$.
Note that we can always set $B=0$ by using a translation in $x$. Therefore, we consider the initial profile
\be
q(x,0)=1-\frac{iA}{\cosh Ax}\,.
\ee
We follow the same steps as in Section \ref{finite_rep}. We have
\be
W_-=W_+=c\mato
0 & 1 \\ 
1 & 0 
\matf\equiv c\,\sigma\,,
\ee
and then,
\be
\Lambda_+(k)=\Lambda_-(k)=\mu(k)\sigma_3\,,
\ee
where $\mu$ satisfies 
\be
\mu^2=k^2-c^2\,.
\ee
For convenience, let us write $c=ig$ (so the unquenched situation corresponds to $g=1$) and 
\be
\mu(k)=\sqrt{k^2+g^2}\,.
\ee
We have
\be
P_-=P_+=\mato
1 & \frac{\mu-k}{g} \\ 
-\frac{\mu-k}{g} & 1 
\matf\equiv P\,.
\ee
As before, let us work with the solution $\Psi^+$ of \eqref{sol_Psi1}-\eqref{sol_Psi2} and set
\be
\Phi^+(x,k)=P^{-1}\Psi^+(x,k)\,.
\ee
After some algebra, we find that $\Phi^+$ is the solution of 
\be
\Phi_x+i\mu(k)\sigma_3\Phi=gQ\Phi
\ee
normalized to $\displaystyle \lim_{x\to\infty}e^{i\mu(k)x\sigma_3}\Phi^+(x,k)=\1$, where
\be
Q(x)=\frac{A}{\cosh Ax}\mato
0 & 1 \\ 
-1 & 0 
\matf
\ee
This is very similar to the rapidly decreasing case with $\mu(k)$ playing the role of $k$. Thus, the effect of the finite density is encapsulated in 
a redefinition of the spectral parameter $k\to\mu(k)$. We simply write $\mu$ in the following.
Setting 
\be
\phi(x,\mu)=e^{i\mu x\sigma_3}\Phi^+\,,
\ee
recall that we want to compute
\be
\mato
b(\mu,c) \\ 
a(\mu,c) 
\matf=\lim_{x\to-\infty}\mato
\phi_{12}(x,\mu) \\ 
\phi_{22}(x,\mu)
\matf\,.
\ee
From here on, the calculations are similar to the rapidly decreasing case. We set 
\be
z(x)=\frac{1}{2}(1-\tanh A x)\,,~~F(z)=\phi_{22}(x,\mu)\,.
\ee
One then finds that $F$ satisfies the hypergeometric differential equation
\be
z(1-z)F^{''}+\left(\frac{1}{2}-i\frac{\mu}{A}-z\right)F^{'}+g^2F=0\,,
\ee
with $\displaystyle \lim_{z\to 0}F(z)=1$. The second boundary condition comes from the condition on $\phi_{12}$ as $x\to\infty$ and reads
\be
\lim_{z\to 0}z^{\frac{1}{2}-\frac{i\mu}{A}}\,F^{'}(z)=0\,.
\ee
Thus,
\be
F(z)=\mathbin{_2F_1}\left(g,-g,\frac{1}{2}-\frac{i\mu}{A};z\right)\,,
\ee
and, after simplification (recall $c=ig$ here),
\bea
a(\mu,c)&=&\frac{\Gamma\left(\frac{1}{2}-\frac{i\mu}{A} \right)^2}
{\Gamma\left(\frac{1}{2}-g-\frac{i\mu}{A} \right)\Gamma\left(\frac{1}{2}+g-\frac{i\mu}{A} \right)}\,,\\
b(\mu,c)&=&-\frac{\sin(\pi g)}{ g }\frac{1}{\cosh\frac{\pi\mu}{A}}\,,
\eea
where $\Gamma$ is the Gamma function. When $g=1$, we find the scattering data corresponding to the initial one-soliton solution 
of the finite density, focusing NLSE equation 
\bea
b(\mu,i)=0~~,~~a(\mu,i)&=&\frac{\mu-\mu_1}{\mu-\mu_1^*}~~,~~\mu_1=\frac{iA}{2}\,.
\eea
Therefore, the quench map $\cQ_{c,i}=\cQ_c$ reads in this case
\bea
\begin{cases}
\frac{\mu-\mu_1}{\mu-\mu_1^*}\,,\\
0 
\end{cases}
\mapsto 
\begin{cases}
\frac{\Gamma\left(1-\frac{i}{A}(\mu-\mu_1) \right)^2}
{\Gamma\left(1-g-\frac{i}{A}(\mu-\mu_1) \right)\Gamma\left(1+g-\frac{i}{A}(\mu-\mu_1)\right)}\,,\\
\frac{\sin(\pi g)}{ g }\frac{i}{\sinh\frac{\pi}{A}\left(\mu-\mu_1 \right)}
\end{cases}
\eea
It looks deceptively identical to the rapidly decreasing case but one has to remember that $\mu$ is a nontrivial function of the original spectral parameter $k$. 
The subtleties of the finite density case are contained in the more complicated analytical properties of the scattering data. That being said, the structural form 
of the scattering coefficients allows one to analyse the focusing to focusing, or focusing to defocusing quench along the same lines as the previous two examples, with 
the same qualitative conclusions.

\subsection{Remarks on the examples}

These three examples show striking similarities in the tools employed. In particular, the mapping of the problem to the hypergeometric differential equation 
is a common feature. One might argue that this is because we have picked a particular soliton solution as initial 
condition in each scenario. However, the close relationship between soliton theory and connection problems for special functions is well-known. The importance 
of special functions in the algebro-geometric approach to the inverse scattering method \cite{Nov,Dub,IM,BBEIM} is also well-known and gives hope that one could develop a 
more systematic theory for the direct part of the inverse scattering method, with applications to a wider class of quench problems. 

In the following Section, we leave the arena of specific examples and consider the general problem of characterising the quench map via its spectral analysis.

\section{The quench map from the ISM: a general approach}\label{sectgen}

In this Section, we address the question of finding  general analytic properties of the quench map by using the full power of the inverse scattering method. For 
simplicity, we consider only the rapidly decreasing focusing case.
Recall that we have $W(x,t)\to 0$ as $|x|\to\infty$. It is convenient to perform a gauge transformation $\Phi:=e^{ik\si_3 x}\Psi$. The transformed matrix potential is then
\beq
	U(x):=e^{ik\si_3 x} W(x,0) e^{-ik\si_3 x}= c \mato 0 & e^{2ikx} q(x,0) \\ -e^{-2ikx}  q^*(x,0) & 0 \matf,
\eeq
where for lightness of notation we keep the $k$ dependence implicit. Note that, for $k$ real, $U^\dag(x)=-U(x)$ and $\Tr(U(x))=0$; that is, $iU$ is in the fundamental representation of the $\frak{su}(2)$ algebra. The fundamental solutions $\Phi_\pm=\Phi_\pm(x)$ to
\beq
\label{Jost_solutions}
	\p_x \Phi_\pm = U \Phi_\pm,\quad \Phi_\pm(x\to\pm\infty,k)=\1_2
\eeq
are $SU(2)$ matrices for $k$ real. Using the convention of ordering the integrals from right (bottom limit) to left (top limit), the solutions can be written as path-ordered integrals:
\beq
	\Phi_-(x) = \Pexp\int_{-\infty}^x dy\,U(y),\quad
	\Phi_+(x) = \Pexp\int_{\infty}^x dy\,U(y).
\eeq
The scattering matrix $S=S(k)$ introduced in \eqref{compute_S} can be written as
\beq
	S = \Phi_-^{\dag}(x)\Phi_+(x) = \Pexp\int_{\infty}^{-\infty} dy\,
	U(y)\label{fact0}
\eeq
(for any $x$). We remark that by standard arguments (which we recall below), one may show that the fundamental solutions $\Phi_\pm$, as functions of $k$, have the following analytic properties:
\beq
	\mato
	(\Phi_-)_{11} & (\Phi_+)_{12}
	\\ (\Phi_-)_{21}  & 
	 (\Phi_+)_{22}\matf \mbox{ is analytic for ${\rm Im}(k)>0$},\quad
	\mato
	(\Phi_+)_{11} & (\Phi_-)_{12}
	\\ (\Phi_+)_{21}  & 
	 (\Phi_-)_{22}\matf \mbox{ is analytic for ${\rm Im}(k)<0$}
	 \label{fact0p}
\eeq
for any $x\in\RR$.

From the discussion in Section \ref{sectqm}, the map ${\cal F}_c$ representing the direct problem with coupling $c$, which maps from the space of initial conditions $q(x)$ to the space of scattering data $S(k)$, is
\beq
	{\cal F}_c: q \mapsto S = 
	 \Pexp \int_{\infty}^{-\infty} dy\,U(y).
\eeq
We are interested in the quench map ${\cal Q}_{c',c}= {\cal F}_{c'} \circ {\cal F}_{c}^{-1}$. We recall that this acts on the space of scattering data and represents the change of scattering data associated to a change of coupling parameter $c\mapsto c'$, keeping fixed the initial condition.

Recall the notation from \eqref{Sab}, with in particular $a=S_{22}$. We assume that $a(k)$ may only have finite-order zeroes in the upper half $k$-plane\footnote{The 
vast majority of the literature on solitons considers only simple zeros but we do not need such a restriction for our arguments in this Section.}. We will show the following:
\begin{proposition}\label{prop} There exist smooth matrix functions $\Theta_\ep(x)$, $\ep\in\{\pm\}$, that also depend on $k$, in the fundamental representation of the $SU(2)$ group for $x,k\in\RR$, such that
\beq
{\cal Q}_{c',c}: S\mapsto S' = \Theta_{-}^\dag(x) S \Theta_{+}(x)
	\label{fact1},
\eeq
where the right-hand side is independent of $x\in\RR$. The following analytic properties hold in the $k$-plane:
\beq\label{Mana}
	M(x):=\frc1{{a}}\mato
	(S^\dag \Theta_-)_{11} &  e^{-2ikx} (\Theta_+)_{12}
	\\ e^{2ikx} (\Theta_-)_{21} &
	 (S\Theta_+)_{22}\matf \ \mbox{is analytic in $k$ for ${\rm Im}(k)>0$}
\eeq
except possibly for isolated singularities at the zeroes of $a$. The following asymptotic conditions in $k$ and in $x$ hold:
\beq\label{asymcond}
	\lim_{k\to i\infty} M(x) = \1_2,\quad \lim_{x\to\infty} \Theta_+(x) = \lim_{x\to-\infty} \Theta_-(x)=\1_2,\quad
	\lim_{x\to-\infty} S \Theta_+(x) = \lim_{x\to\infty} \Theta_-^\dag(x) S.
\eeq
Finally, let $m(x)$ be a column vector of $M(x)$, and let $k=k_0$ be a zero of $a(k)$ in the upper half $k$-plane of order $j\geq 1$. If we assume that $m(x)$ has a finite-order pole at $k=k_0$ of order $\ell\geq0$ (or no pole, $\ell=0$), then there exists $b_0\in\CC$ such that $m(x)$ has the form
\beq\label{singula}
	m(x) \propto(k-k_0)^{-\ell}\mato -b_0 \\ 1 \matf + O\lt((k-k_0)^{1-\ell}\rt).
\eeq
If the coefficient $b(k)$ in \eqref{Sab} can be extended analytically to $k_0$, then $b_0 = b(k_0)$.
\end{proposition}

In Appendix \ref{app} we provide a more complete analysis of the form of $m(x)$ in the vicinity of $k=k_0$, see \eqref{thetaTaylor} and \eqref{ttheta}.

With an explicit parametrization
\beq\label{paramT}
	\Theta_\pm = \mato  u^*_\pm & v_\pm \\ - v^*_\pm & u_\pm \matf,\quad |u_\pm|^2 + |v_\pm|^2 = 1,
\eeq
the analyticity conditions \eqref{Mana} and the asymptotic conditions in $k$ translate into the statement that
\beq\label{fghj}
	 u^*_-(x) + \frc{b}a v^*_-(x),\quad 
	-e^{2ikx} \frc1a  v^*_-(x),\quad e^{-2ikx} \frc1a v_+(x),\quad
	 u_+(x) - \frc{ b^*}a v_+(x)
\eeq
are analytic for ${\rm Im}(k)>0$ except for possible isolated singularities at the zeroes of $a(k)$, and tend to 1, 0, 0 and 1 respectively as $k\to i\infty$.

Let us fix $x=0$. The above proposition can then be stated as follows: there exists four functions $f$, $g$, $h$ and $j$ of the variable $k$  analytic in the upper half plane except for possible isolated singularities at the zeroes of $a$,  whose limits as $k\to i\infty$ are 1, 0, 0 and 1 respectively, and with the properties that
\beq\label{fghj1}
	f f^* + g g^* + g f^*\,b + f  g^*\, b^* = 1, \quad h  h^* + j j^* + j  h^*\,b + h  j^* \, b^*  = 1,
\eeq
and that the possible singularities of the column vectors $(f,g)^T$ and $(h,j)^T$, at the zeroes $k_0$ of $a$, are described as in \eqref{singula}, such that the scattering data after the quench are given by
\beq\label{apbp}
	a' = (fj-gh)a,\quad b' = j f^* \,b + h g^* \, b^* + j g^* +h f^*.
\eeq
The functions $f,g,h,j$ are the matrix elements of $M(0) = \lt(\begin{matrix} f & h \\ g & j\end{matrix}\rt)$, equivalently the four quantities in \eqref{fghj}, respectively, at $x=0$.

\noindent {\em Sketch of the proof.} We are looking for an expression for $S' = \mathcal{Q}_{c',c}(S)$. We will use the notation $\delta c=c'-c$. Consider the solutions $\Theta_{\ep}(x)$, for $\ep\in\{\pm\}$, to the scattering equations
\beq\label{tu}
	\p_x \Theta_{\ep}(x) = \frc{\delta c}c U_\ep(x) \Theta_{\ep}(x)
\eeq
with potential
\beq
	U_\ep(x) = \Phi_\ep^{\dag}(x) U(x) \Phi_\ep(x),
\eeq
normalized as $\Theta_{\ep}(x\to\ep\infty)=\1_2$. Here and below, except when explicitly displayed or stated, $U(x)$ and $\Phi_\ep(x)$ are evaluated at $c$.

Note that $U(x,c') = (c'/c) \,U(x,c)$. Using this simple relation along with \eqref{Jost_solutions} and the uniqueness of the initial value problems, we deduce that
\beq
	\Theta_{-}(x) = \Phi_-^\dag(x,c) \Phi_-(x,c'),\quad
	\Theta_{+}(x) = \Phi_+^\dag(x,c) \Phi_+(x,c').
\eeq
Then, from \eqref{fact0} we remark that
\beq\label{tst}
	\Theta_{-}^\dag(x)\, S\, \Theta_{+}(x) = \Phi_-^\dag(x,c')\Phi_+(x,c') = S'.
\eeq
Setting $\Theta_\ep = \Theta_{\ep}(x)$ for any $x\in\RR$, we have \eqref{fact1}. We now use the scattering problems \eqref{tu} in order to deduce the analytic properties \eqref{Mana} of $\Theta_\ep$ as functions of $k$.

Recall that one can solve recursively the scattering equation \eqref{Jost_solutions}, with appropriate asymptotic conditions, by re-writing them in integral form,
\beq
	\Phi_\ep(x) = \1_2 + \int_{\ep\infty}^x dy\,U(y)\Phi_\ep(y).
\eeq
Let us make the transformation:
\beq\label{tr}
	\Phi_\ep(x) = e^{ik\sigma_3 x}\t\Phi_\ep(x) e^{-ik\sigma_3 x}
\eeq
so that
\beq\label{tphi}
	\t\Phi_\ep(x) = \1_2 + \int_{\ep\infty}^x dy\,e^{ik\sigma_3(y-x)} W(y)
	\t\Phi_\ep(y) e^{-ik\sigma_3(y-x)}.
\eeq
Let us define the column vectors $\frak{f}_{\ep\eta}$, for $\eta=\pm$, by $\t\Phi_\ep = (\frak{f}_{\ep+},\frak{f}_{\ep-})$. Applying, to the right of \eqref{tphi}, the projectors $\PP_\eta$ onto the eigenvectors $v_\eta$ of $\sigma_3$ with eigenvalues $\eta$, we conclude that
\beq
	\frak{f}_{\ep\eta}(x) = v_\eta
	+ \int_{\ep\infty}^x dy\,e^{ik(\sigma_3-\eta)(y-x)} W(y)
	\frak{f}_{\ep\eta}(y).
\eeq
By standard arguments, recursively solving this integral equation, one deduces the following: $\frak{f}_{\ep,-\ep}(x)$ are analytic in $k$ for ${\rm Im}(k)>0$, and tend to $v_{-\ep}$ as $k\to +i\infty$; and $\frak{f}_{\ep\ep}(x)$ are analytic in $k$ for ${\rm Im}(k)<0$, and tend to $v_\ep$ as $k\to -i\infty$; these statements being valid uniformly for $x\in\RR$. This in particular implies \eqref{fact0p}. We remark that, from the representation $a=\det(\frak{f}_{-+},\frak{f}_{+-})$ \cite{FT}, one finds that $a$ is analytic in the upper half $k$-plane and $\lim_{k\to i\infty} a(k)=1$.

Similarly, let us re-write equation \eqref{tu}, with appropriate asymptotic conditions, as integral equations,
\beq\label{thetau}
	\Theta_{\ep}(x) = \1_2 + \frc{\delta c}c \int_{\ep\infty}^x dy\,
	U_\ep(y)\Theta_{\ep}(y).
\eeq
Let us introduce the column vectors $f_{\ep\eta}$ by $\Phi_\ep=(f_{\ep+},f_{\ep-})$, and recall that $S = \mato  a^* & b \\ - b^* & a\matf$ (using \eqref{Sab} and $c\in i\RR^+$), with $|a|^2+|b|^2=1$. Let us further define
\beq
	\Phi_> = (f_{-+},f_{+-}),\quad \Phi_< = (f_{++},f_{--}),\quad
	U_> = \Phi_<^\dag U \Phi_>,\quad
	U_< = \Phi_>^\dag U \Phi_<.
\eeq
The relation \eqref{fact0} implies $f_{++}= a^* f_{-+} -  b^* f_{--}$ and $f_{+-} = b f_{-+} + a f_{--}$, and thus we have
\beq
	\Phi_+ = \frc1{a}\,\Phi_> \mato 1 & 0 \\ - b^* & a\matf
	= \frc1{ a^*}\,\Phi_<\mato { a^*} & b \\ 0 & 1\matf,\quad
	\Phi_- = \frc1{a}\,\Phi_> \mato a & -b \\ 0 & 1\matf
	= \frc1{ a^*}\,\Phi_< \mato 1 & 0 \\  b^* &  a^*\matf.
\eeq
From this we remark that
\beq
	U_+ = \frc1{a} J_+^{-1} U_> J_+ = \frc1{ a^*} J_+^\dag
	U_< J_+^{\dag\,-1},\quad
	U_- = \frc1{a} J_-^{-1} U_> J_-=\frc1{ a^*} J_-^\dag
	U_< J_-^{\dag\,-1}
\eeq
where
\beq\label{jpm}
	J_+ = \mato 1 & 0 \\ - b^* & a\matf = \PP_+ + \PP_- S,\quad
	J_- = \mato a & - b \\ 0 & 1\matf = \PP_- + \PP_+ S^\dag.
\eeq
We now define
\beq
	\Theta_{\ep>}(x) = J_\ep \Theta_\ep(x) J_\ep^{-1},\quad
	\Theta_{\ep<}(x) = J_\ep^{\dag\,-1} \Theta_\ep(x)  J_\ep^{\dag}.
\eeq
Applying these similarity transformations to the equation \eqref{thetau}, we obtain
\beq\label{tek}
	\Theta_{\ep\kappa}(x) = \1_2 + \frc{\delta c}c \frc1{a^\kappa} \int_{\ep\infty}^x dy\,
	U_\kappa(y)\Theta_{\ep\kappa}(y)
\eeq
where $\ep\in\{\pm\}$ and $\kappa\in\{>,<\}$ and where $a^> = a$ and $a^< =  a^*$.

Let us make the transformation $\Theta_{\ep\kappa}(x) = e^{ik\sigma_3x}\t\Theta_{\ep\kappa}(x) e^{-ik\sigma_3 x}$ and $\Phi_\kappa(x) = e^{ik\sigma_3x} \t\Phi_\kappa(x) e^{-ik\sigma_3 x}$. We then obtain
\beq\label{thetau2}
	\t\Theta_{\ep\kappa}(x) = \1_2 + \frc1{a^\kappa}
	\int_{\ep\infty}^x dy\,\frc{\delta c}c \,
	e^{ik\sigma_3(y-x)}\, \t\Phi_{-\kappa}^\dag(y) W(y)
	\t\Phi_\kappa(y)\,
	\t\Theta_{\ep\kappa}(y)\, e^{-ik\sigma_3(y-x)}
\eeq
(where here $-\kappa$ is $<$ ($>$) if $\kappa$ is $>$ ($<$)).
We remark that $\t \Phi_>(x) = (\frak{f}_{-+}(x),\frak{f}_{+-}(x))$ is analytic in $k$ for ${\rm Im}(k)>0$ and tend to $\1_2$ as $k\to i\infty$, and that $\t \Phi_<(x) = (\frak{f}_{++}(x),\frak{f}_{--}(x))$ is analytic in $k$ for ${\rm Im}(k)<0$ and tend to $\1_2$ as $k\to -i\infty$ (both uniformly in $x$). Therefore, by recursively solving the integral equation and using the fact that $a(k)$ tends to 1 as $k\to i\infty$, we conclude that: $\t\Theta_{+>}(x)\PP_-$ and $\t\Theta_{->}(x)\PP_+$ are both analytic in $k$ for ${\rm Im}(k)>0$, except possibly at the positions of the zeroes of $a=a(k)$ in the upper-half plane, and tend to $\PP_-$ and $\PP_+$, respectively, as $k\to i\infty$; and that $\t\Theta_{+<}(x)\PP_+$ and $\t\Theta_{-<}(x)\PP_-$ are both analytic in $k$ for ${\rm Im}(k)<0$, except possibly at the positions of the zeroes of 
$ a^*=  a^*(k)$ in the lower-half plane, and tend to $\PP_+$ and $\PP_-$, respectively, as $k\to -i\infty$. Clearly, the same analyticity properties hold for
\beq
	M_>(x):=\sum_{\ep} \t\Theta_{\ep>}(x)\PP_{-\ep}\quad\mbox{and}\quad
	M_<(x):=\sum_\ep \t\Theta_{\ep<}(x)\PP_\ep,
\eeq
respectively, with the same asymptotic behaviours as $k\to\pm i \infty$. Using the relations
\beq
	J_+^{-1} = \frc1a(\PP_- + S^\dag\PP_+),\quad J_-^{-1} = \frc1a(\PP_+ + S\PP_-)
\eeq
along with \eqref{jpm} one obtain the explicit forms
\begin{align}
	& M_>=\frc1{{a}}\mato
	(S^\dag \Theta_-)_{11} & e^{-2ikx} (\Theta_+)_{12}
	\\ e^{2ikx} (\Theta_-)_{21} &
	 (S\Theta_+)_{22}\matf 
	 \n
	&M_<=\frc1{{ a^*}}
	\mato
	(S \Theta_+)_{11} & e^{-2ikx}  (\Theta_-)_{12}
	\\ e^{2ikx}  (\Theta_+)_{21}  & 
	 (S^\dag \Theta_-)_{22}\matf.
\end{align}
With the parametrization \eqref{paramT}, one finds that
\beq\label{Mpm}
	M_< = \sigma_2  M^*_> \sigma_2.
\eeq
Therefore, the analyticity and asymptotic conditions on $M_>$ and $M_<$ are equivalent, and it is sufficient to consider only one group of conditions, leading to \eqref{Mana} with $M=M_>$. Finally, the asymptotic conditions \eqref{asymcond} are consequences of the above discussion, in particular the last one in \eqref{asymcond} coming from the consistency of the representation \eqref{fact1} along with the normalizations $\Theta_+(\infty)=\Theta_-(-\infty)=\1_2$.

Finally, let us analyze the singularities of $M(x)$ at the positions of the zeroes of $a$. Recursively solving \eqref{thetau2} (with $\kappa$ being $>$) one would, {\em a priori}, find higher and higher powers of the poles of $1/a$, and thus one could expect essential singularities to appear in $M(x)$. However, from the representation $a=\det(\t\Phi_>(x))$, it is clear that at a zero $k=k_0$ of $a$ in the upper half plane, the matrix $\t\Phi_>(x)$ has an eigenvector with eigenvalue 0, that is $\frak{f}_{+-}(x) = b_0\frak{f}_{-+}(x)$ for some $b_0\in\CC$ independent of $x$ \footnote{Note that if the function $b(k)$ appearing in $S(k)$ \eqref{Sab} can be analytically continued to $k_0$ (this is possible in several situations, e.g. compactly supported or exponentially decaying $q(x)$), then $b_0=b(k_0)$ as per \eqref{fact0}; but the argument we present does not rely on this.}. Therefore $\Phi_>$ has the zero right-eigenvector $(-b_0,1)^T$, and if a column vector $m$ of $M$ has a pole of order $\ell\geq0$ (or has no pole, $\ell=0$) at $k=k_0$, then we must have $m \propto (k-k_0)^{-\ell}\, (-b_0,1)^T$ for some nonzero proportionality constant. Here $\ell$ may depend on the zero of $a$ and on the column vector that we are looking at. A more precise analysis gives \eqref{singula}.
\eproof
\begin{remark} Equations \eqref{tu} represent two different scattering problems, with potentials $U_+$ and $U_-$, each of which has two independent fundamental solution. One may then consider the four $SU(2)$-valued functions $\Theta_{\ep\eta}$, for $\ep,\eta\in\{\pm\}$,
\beq\label{tu2}
	\p_x \Theta_{\ep\eta}(x) = \frc{\delta c}c U_\ep(x) \Theta_{\ep\eta}(x),
\eeq
normalized as $\Theta_{\ep\eta}(x\to\eta\infty)=\1_2$. Above we only used $\Theta_{\ep}(x) = \Theta_{\ep\ep}(x)$, and a similar analysis applied to $\Theta_{\ep,-\ep}(x)$ does not provide new information. The associated scattering matrices are
\beq
	Q_{\ep}(c,c') = \Theta_{\ep,-}^{-1}(x)\Theta_{\ep,+}(x)
\eeq
independently of $x$. Using \eqref{fact0}, we notice that $U_-(x) = S U_+(x) S^\dag$, whereby $\Theta_{--}(X) = S \Theta_{+-}(x)S^\dag$ and $\Theta_{-+}(x) = S \Theta_{++}(x)S^\dag$. With \eqref{tst}, this gives the representation of the scattering matrices $Q_\ep$ in terms of $S$ and $S'$:
\[
	Q_-(c,c') = S'S^{-1},\quad Q_+(c,c') = S^{-1}S'.
\]
That is, this ``higher-level'' scattering problem has scattering matrix equal to the change of the ``low-level'' scattering matrix under the quench. Path-ordered representations can also be obtained as usual,
\beqa
	Q_-(c,c') &=& \Pexp \int_{\infty}^{-\infty}dx\, \frc{\delta c}c\,U_-(x)\n
	Q_+(c,c') &=&\Pexp \int_{\infty}^{-\infty}dx\, \frc{\delta c}c\,U_+(x).\eeqa
\end{remark}

Proposition \ref{prop} gives a set of analytic requirements for objects from which the quench map ${\cal Q}_{c',c}$ is constructed. For instance, the solution $\Theta_- = \Theta_+ = \1_2$ gives the case $c'=c$. For free initial data $c=0$, the scattering matrix $S$ is the identity $\1_2$, and we recover the factorization \eqref{fact0} with \eqref{fact0p} upon identifying $\Theta_\pm = \Phi_\pm$. However, this set of requirements does not uniquely fix the quench map. In particular, there is no information, in the analytic requirements, about the initial and final coupling strengths $c$ and $c'$. 

We may ask if the requirements of Proposition \ref{prop} give nontrivial constraints. A naive counting argument is as follows, omitting for now the information about the singularities at the zeroes of $a$: the four complex functions $f$, $g$, $h$ and $j$ account for eight real smooth functions, and analyticity in the upper half plane effectively reduces this number to four. The two relations \eqref{fghj1} further reduce this number to two. Thus, given $a$ and $b$, the relation \eqref{apbp} provides a two-real-dimensional space of solutions. The space of scattering data is also two-real-dimensional, as, 
besides the structure of zeroes of $a(k)$, it can be reconstructed from the complex reflection coefficient $b(k)/a(k)$; and therefore it seems as though 
the requirements do not determine the post-quench scattering data. However, the space of effective free real functions is reduced by the gauge freedom 
afforded by \eqref{fact1}: $\Theta_\pm \mapsto C\Theta_\pm$ for any $C\in SU(2)$ in the centralizer of $S$, and $\Theta_\pm \mapsto \Theta_\pm C$ for 
$C\in SU(2)$ in the centralizer of $S'$ (assuming, for instance, that $S$ or $S'$ lies in a subset of $SU(2)$ with nontrivial centralizer). 
Further, the isolated singularities at the positions of the zeroes of $a$ in the upper half $k$-plane give additional constraints. 
In Appendix \ref{app} we provide an analysis of these constraints. Finally, combining equations \eqref{apbp} with analytical requirements for the post-quench 
coefficient $a'(k)$ might lead to additional information.  A more complete analysis may be useful, and we hope to come back to these issues in a future work.

\section{Dual quench, B\"acklund transformations, GLM equations and Rosales expansion}\label{real_quench}

The quench problem of Figure \ref{quench_scattering} corresponds to an abrupt change of the coupling parameter $c$ of the NLSE. This is a quench with respect to the field $q(x)$: the field $q(x)$ is kept the same, but the scattering data $S(k)$ is abruptly changed (via the quench map) because the scattering transform, associated to the NLSE, is modified. As we explained, the evaluation of the quench map is the classical analogue of the quantum overlap between eigenstates associated to different Hamiltonians.

Another natural protocol is a slow, adiabatic change of the coupling parameter $c$. In the quantum realm, under adiabatic changes the eigenstates of the initial Hamiltonian are naturally mapped to those of the final Hamiltonian, whereby all natural conserved quantities (the energy, the number of particles, the asymptotic momenta, etc.) are unchanged. In the classical re-interpretation, the scattering data play the role of these conserved quantities. Hence, one considers the {\em dual quench} protocol: a quench with respect to the scattering data. The scattering data $S(k)$, associated to a field $q(x)$ for the NLSE with coupling $c$, is kept fixed, and we ask about the field $\tilde{q}(x)$ having the same scattering data $S(k)$ but under the NLSE with coupling $c_0$. This is represented on Figure \ref{quench_real1} and the map $\cB$ captures the effect of this dual quench protocol.
\begin{figure}[htp]
\def\Id{{\rm id}}
\begin{diagram}
q(x) & & \rTo^{\cF_c} & & S(k)  \\
\dTo^{\cB} & &  & & \dTo_{\Id} \\
\tilde{q}(x) & &  \rTo^{\cF_{c_0}} & & S(k) \\
\end{diagram}
\caption{The dual quench, keeping the scattering data fixed.}
\label{quench_real1}
\end{figure}

Following the idea coming from the quantum picture, we would expect the map $\cB$ to represent the result, in the limit $t\to\infty$, of the adiabatic protocol whereby $q(x)$ is first {\em backward} evolved until time $-t$ under the NLSE with coupling $c$, then {\em forward} evolved back to time $0$ under the NLSE with a slowly varying coupling from $c$ (at time $-t)$ to $c_0$ (at time 0). This is the classical analogue of the quantum scattering map for this adiabatic process, $ \lim_{t\to\infty} \Pexp\lt[ -i\int_{-t}^0ds\, H_{c(s)}\rt] e^{iH_ct}$ with $c(s) = (-s c + (s+t)c_0)/t$.

It is well known that B\"acklund transformations play a distinguished role in the theory of integrable PDEs and it is natural 
to ask whether or not the map $\cB$ can correspond to a certain B\"acklund transformation (BT) taking $q(x)$ to $\tilde{q}(x)$. 
It is natural to reinstate the time dependence as B\"acklund transformations map solutions of a PDE to solutions of a (possibly different) PDE. 
This question then amounts to showing that the map  $\cB$ on Figure \ref{quench_real1} can be seen as the $t=0$ restriction of a B\"acklund transformation (BT), 
see Figure \ref{quench_real2}.
\begin{figure}[htp]
\def\Id{{\rm id}}
\begin{diagram}
q(x,t) & & \rTo{t=0} & & q(x,0) & & \rTo^{\cF_c} & & S(k)  \\
\dTo^{BT} & &  & & \dTo^{\cB} & &  & & \dTo_{\Id} \\
\tilde{q}(x,t) & &  \rTo^{t=0} & & \tilde{q}(x,0) & &  \rTo^{\cF_{c_0}} & & S(k) \\
\end{diagram}
\caption{The dual quench as a B\"acklund transformation.}
\label{quench_real2}
\end{figure}

If always true for any value of the quench, it would imply that quenches with respect to the scattering data are of a very special nature: 
all the resulting fields $\tilde{q}(x)$ are in the orbit of the initial condition $q(x)$ under the action of the group of 
B\"acklund transformations\footnote{Indeed, it is known that B\"acklund transformations form a group acting on solutions of integrable PDEs, see e.g. \cite{Sem}.}.
In that case, it would be interesting to obtain an explicit characterization of these transformations in 
terms of known BT's and if possible, a formula relating $\tilde{q}(x,t)$ to $q(x,t)$. 
Note that the crux of the matter here is not to find $\tilde{q}$ since the inverse scattering method tells us how to do this, by applying 
the inverse transform $\cF^{-1}_{c_0}$ to the scattering data $S(k)$. 
The question is whether or not the obtained solution $\tilde{q}(x,t)$ can always be seen as an explicit BT of $q(x,t)$
for all $t\ge 0$, in which case the question is answered in particular at $t=0$.

Beyond its interest in terms of adiabatic changes, this question may also shed light on the original quench problem. Indeed, the map $\cB$ and the quench map $\cQ_{c,c_0}$ are related as follows:
\be\label{intertw}
\cF_{c_0}\,\cB=\cQ_{c,c_0}\,\cF_{c_0}\,.
\ee
This is an intertwining equation where the intertwiner is the direct transform $\cF_{c_0}$ of the inverse scattering method.
Hence, knowing $\cB$ provides us with a formal representation of the quench map as
\be
\cQ_c=\cF_{c_0}\,\cB\,\cF_{c_0}^{-1}\,.
\ee

In general, in the NLSE, we note that the interpretation of $\cB$ as a BT is either a trivial problem or is very difficult. In the rest of this Section, we present some results towards the answer in 
some difficult cases. As for the trivial answer, remark that in the rapidly decreasing case, for a quench preserving the nature of the 
system \ie $\frac{c}{c_0}\in \RR$, then $\cB$ is simply a rescaling 
\be
\tilde{q}(x,0)=\frac{c}{c_0}q(x,0)\,,
\ee
which is compatible with the time evolution. So indeed, $\cB$ can be seen as the $t=0$ restriction of the simplest BT 
\be
\tilde{q}(x,t)=\frac{c}{c_0}q(x,t)\,.
\ee
In all other cases, this simple answer does not hold. Indeed, in the focusing to defocusing case (or vice versa), $\frac{c}{c_0}\in i\RR$ 
and the previous rescaling is not compatible with the time evolution prescribed by NLSE. To see this, note that if $c=i \beta$, $c_0=\beta_0$ with
$\beta,\beta_0\in \RR$ then
\be
i\partial_t \tilde{q}+\partial_x^2 \tilde{q}-2c_0^2|\tilde{q}|^2\tilde{q}=0\,,
\ee
becomes
\be
i\partial_t q+\partial_x^2 q-2\beta^2|q|^2q=0\,,
\ee
while it should be $i\partial_t q+\partial_x^2 q+2\beta^2|q|^2q=0$. In the finite density case, things are even more complicated. Even for a quench preserving the focusing or defocusing nature of the equation of motion, 
the rescaling alters the boundary conditions and so is not compatible with the prescribed system.

Remark that for integer values of the quench and for a corresponding very specific choice of initial conditions, 
we see from the three explicit examples treated previously that the net effect on the scattering data is simply to change the number of zeros 
in the spectral function $a(k)$, while maintaining its rational form and the reflectionless property $b(k)=0$. Using \ref{intertw}, this translates into a map $\cB$ whose effect is to change accordingly the fixed-$c$ scattering data associated to $q(x)$. As we recall below in our short review on BTs, such an effect is easily implemented by using a special type of BT that can add (or remove) 
exactly one zero in the spectral function $a(k)$ and keep $b(k)=0$. 
These BTs are well-known and their ``real'' counterpart on $q(x,t)$ can be written explicitely. Combined with the observation that 
generic scattering data are completely determined by these zeros and the radiative part corresponding to the continuous spectrum of the auxiliary problem, 
this gives us a strategy to characterize $\tilde{q}$ in terms of $q$ via a set of special BTs and a transformation taking care of the radiative part. 
It turns out that the latter is well encoded in the so-called Rosales solution of NLSE \cite{Ros} which has a simple interpretation in 
terms of the usual Gelfand-Levitan-Marchenko approach to the inverse part of the inverse scattering method. 
In the rest of this Section, we present the aforementioned two ingredients (special BT's and Rosales solution) and then combine them to 
obtain some more understanding of the map $\cB$ in the generic case.

\subsection{Review of B\"acklund transformations }\label{review_BT}

The general notion of B\"acklund transformations in integrable PDEs, that is of a transformation $q\mapsto \tilde{q}$ such that $\tilde{q}$ 
is a solution of PDE$2$ if $q$ is a solution of PDE$1$ and vice versa, is closely related to the theory of the so-called Darboux matrices which acts on 
the wavefunctions of the associated auxiliary problem encoding the involved PDEs. That is why in the context of integrable PDEs, they are sometimes called 
Darboux-B\"acklund transformations. The theory is well-known and we follow here the book \cite{GHZ}. 
We are interested in the case where PDE$1$ and PDE$2$ are the 
same equation \ie the NLSE in our case. We focus on the rapidly decreasing case here to present the ideas and comment on an extension 
to the finite density case at the end of this Section. Let us recall the main steps to obtain 
a transformation that precisely adds or remove one simple zero to the scattering coefficient $a(k)$ while keeping $b(k)$ unchanged.

The idea is to construct a matrix $D$ such that $\widetilde{\Psi}=D\,\Psi\,D_0$ is solution of 
\bea
\begin{cases}
\widetilde{\Psi}_x=(-ik\sigma_3+\widetilde{W}(x,t))\widetilde{\Psi}\,,\\
\widetilde{\Psi}_t=(-2ik^2\sigma_3+\widetilde{P}(x,t,k))\widetilde{\Psi}\,,
\end{cases}
\eea
when $\Psi$ is solution of
\bea
\begin{cases}
\label{bis1}
\Psi_x=(-ik\sigma_3+W(x,t))\Psi\,,\\
\Psi_t=(-2ik^2\sigma_3+P(x,t,k))\Psi\,,
\end{cases}
\eea
as before. The constant matrix $D_0$ is used to preserve the normalisation of $\Psi$. One immediately obtains the equations
\bea
&&(-ik\sigma_3+\widetilde{W})D=D(-ik\sigma_3+W)+D_x\,,\\
&&(-2ik^2\sigma_3+\widetilde{P})D=D(-2ik^2\sigma_3+P)+D_t\,.
\eea
Let us choose 
\be
D(x,t,k)=k\1-\Sigma(x,t)\,.
\ee
Then, the new potential $\widetilde{W}(x,t)$ is readily obtained from the old one $W(x,t)$ by
\be
\label{change_potential}
\widetilde{W}=W+i[\sigma_3,\Sigma]\,.
\ee
The effect on the scattering data is given by
\be
\label{effect}
\widetilde{S}(k)=(D^-)^{-1}\,S(k)\,D^+
\ee
where the new scattering matrix is defined by
\be
\widetilde{\Psi}^+(x,k)=\widetilde{\Psi}^-(x,k)\,\widetilde{S}(k)\,,
\ee
and
\be
D^\pm=\left(\lim_{x\to\pm\infty}D\right)^{-1}\,.
\ee
It remains to find an appropriate $\Sigma$. It is shown in \cite{GHZ} that one can choose
\be
\Sigma=H\Lambda H^{-1}\,,~~\Lambda=\left(\begin{array}{cc}
k_0 &0\\
0 & k_0^*
\end{array}\right)\,,
\ee
where $\text{Im}~ k_0>0$ and $H$ is made of two column vectors that are solution of \eqref{bis1} for $k=k_0$ and $k=k_0^*$. It can be chosen as
\be
H=\left(\begin{array}{cc}
\alpha(x,t,k_0) & \beta^*(x,t,k_0)\\
\frac{c}{c^*}\beta(x,t,k_0) & \alpha^*(x,t,k_0)
\end{array}\right)\,,
\ee
where $\left(\begin{array}{c}
\alpha(x,t,k_0) \\
\frac{c}{c^*}\beta(x,t,k_0) \end{array}\right)$ 
is solution of \eqref{bis1} for $k=k_0$.
In this case, we obtain
\be
\Sigma=\frac{1}{1-\frac{c}{c^*}|\sigma|^2}\left(\begin{array}{cc}
k_0-k_0^*\frac{c}{c^*}|\sigma|^2 &(k_0^*-k_0)\sigma^*\\
\epsilon(k_0-k_0^*)\sigma & k_0^*-k_0\frac{c}{c^*}|\sigma|^2
\end{array}\right),\,\sigma=\frac{\beta}{\alpha}\,.
\ee
Let us denote 
\be
\label{columns}
\Psi^-=(R,\widehat{R})e^{-ikx\sigma_3}\,,~~\Psi^+=(\widehat{L},L)e^{-ikx\sigma_3}\,.
\ee
Given the analytic properties of $\Psi^\pm$, a candidate for $\alpha$ and $\beta$ is 
\be
\left(\begin{array}{c}
\alpha(x,t,k_0) \\
\frac{c}{c^*}\beta(x,t,k_0) \end{array}\right)=Re^{-ik_0x}-\mu Le^{ik_0x}\,,
\ee
for some nonzero $\mu$. Then it can be shown the following \cite{GHZ}:

\begin{enumerate}
\item If $k_0$ is not a zero of $a(k)$ then $\displaystyle\lim_{x\to\infty}\sigma=0$ and $\displaystyle\lim_{x\to-\infty}\sigma=\infty$ so that 
\be
D^+=\left(\begin{array}{cc}
\frac{1}{k-k_0} &0\\
0 & \frac{1}{k-k_0^*}
\end{array}\right)\,,~~D^-=\left(\begin{array}{cc}
\frac{1}{k-k_0^*} &0\\
0 & \frac{1}{k-k_0}
\end{array}\right)\,.
\ee

\item If $k_0$ is already a zero of $a(k)$ and we choose $\mu$ to be different from the corresponding norming constant then 
$\displaystyle\lim_{x\to\infty}\sigma=\infty$ and $\displaystyle\lim_{x\to-\infty}\sigma=0$ so that
\be
D^+=\left(\begin{array}{cc}
\frac{1}{k-k_0^*} &0\\
0 & \frac{1}{k-k_0}
\end{array}\right)\,,~~D^-=\left(\begin{array}{cc}
\frac{1}{k-k_0} &0\\
0 & \frac{1}{k-k_0^*}
\end{array}\right)\,.
\ee
\end{enumerate}

Inserting in \eqref{effect}, we get the effect on the scattering data as desired. In both cases, $b(k)$ is untouched: $\tilde{b}(k)=b(k)$. In the first, 
case $\tilde{a}(k)=\frac{k-k_0}{k-k_0^*}a(k)$ and the net effect of the transformation is to add a simple zero in the upper half-plane to $a(k)$. In the 
second case, $\tilde{a}(k)=\frac{k-k_0^*}{k-k_0}a(k)$, and the net effect is to remove the simple zero $k_0$ of $a(k)$. We can also extract the corresponding 
transformation of the solution $q(x,t)$ in real space, which was our motivation for introducing these transformations in this Section,
\be
\tilde{q}=q-\frac{2i(k_0^*-k_0)}{c(1-\frac{c}{c^*}|\sigma|^2)}\sigma^*
\ee

\subsection{Review of the Rosales solution}\label{review_Rosales}

In \cite{Ros}, Rosales proposed a direct solution method for certain integrable nonlinear PDEs based on a series expansion approach. 
Writing
\begin{eqnarray}
q(x,t)=\sum_{n=1}^\infty \varepsilon^n\psi_n(x,t)\,,
\end{eqnarray}
with $\psi_n\to 0$ as $|x|\to \infty$, and inserting in NLSE, Rosales reduces the problem to finding
a function $\Phi_n(k_1,\ldots,k_n)$ by making the ansatz
\begin{eqnarray}
\psi_n(x,t)=\int_{\CC^n}\Phi_n(k_1,\ldots,k_n)e^{i\Omega_n}d\lambda(k_1)\ldots
d\lambda(k_n)\,.
\end{eqnarray}
Here $d\lambda$ is a measure over $\CC$ and\footnote{Note that we took the convention $-k_jx$ instead of $k_jx$ in $\Omega_n$ in the original paper
\cite{Ros} to be in line with our conventions in the GLM method below.}
\begin{eqnarray}
\Omega_n=\sum_{j=1}^n\left(-k_jx+(-1)^jk_j^2t\right)\,.
\end{eqnarray}
The final results reads
\begin{eqnarray}
q(x,t)=\varepsilon\sum_{n=0}^\infty (-c^2\varepsilon^2)^n\int_{\CC^{2n+1}}\frac{e^{i\Omega_{2n+1}}}
{\prod\limits_{j=1}^{2n}(k_j+k_{j+1})}\prod\limits_{j=0}^n
d\lambda(k_{2j+1}) \prod\limits_{j=1}^nd\mu(k_{2j})\qquad
\end{eqnarray}
with $d\mu^*(-k^*)=d\lambda(k)$. This special relation comes from the fact that both $q$ and $q^*$ appear in NLSE.
Rosales then shows that this series can be resummed for not necessarily small values of $\varepsilon$ and in particular one can set $\varepsilon=1$.
The point is that the entire content of the inverse scattering method can be encapculated in the 
choice of the measure. For instance, by choosing a purely discrete measure, one can reproduce the 
well-known solitons solutions of the inverse scattering method. In the present context, we 
want to make use of the Rosales resolution in the opposite situation where only the continuous spectrum is present. This 
gives us a series representation of the so-called radiative solution of NLSE.  
This corresponds to choosing a continuous measure of the form 
$d\lambda(k)=\beta\rho(k)dk$ for some function $\rho$ with support on $\RR$ and some constant $\beta$. An interesting aspect of Rosales's approach is that one can perform 
a resummation of the series representation. In general, the resummation is rather formal but Rosales gives some conditions under which it is rigorously 
meaningful. It is beyond the scope of this paper to investigate these precise conditions. We will explain below how this gives us a means to 
discuss the issue at hand: the quench from a focusing to a defocusing system (or vice versa). 

We now show how the Rosales expansion is related to the standard Gelfand-Levitan-Marchenko (GLM) approach to the inverse scattering method. Although 
this is known 
to the experts, see \cite{FGZ} for a general theory and also \cite{Vthesis} for a form closer to our needs here, 
we find it useful to recall the main steps in what follows.
The Neumann series approach to the solution of the GLM equation in the case where the scattering data $a(k)$ has no zeros 
produces exactly a Rosales type solution, for a special case of the measure $d\lambda(k)$. 
The interpretation of $\rho(k)$ in this context is clear: it is the reflection coefficient $\frac{b(k)}{a(k)}$. The constant $\beta$ is related to the parameter 
$c$ (see \eqref{link} below).
We recall the main steps of the GLM method in the rapidly decaying case, referring to \cite{FT} for details, and then present the Neumann series solution of the GLM equations 
to make contact with the Rosales solution. 

The starting point of the GLM method 
is to introduce a kernel $K$ via the relation
\begin{eqnarray}
\Psi^-(x,k)=
e^{-ikx\sigma_3}+\int^x_{-\infty}
ds~K(x,s)e^{-iks\sigma_3}\,,
\end{eqnarray}
with
\begin{eqnarray}
K(x,y)=\left(\begin{array}{cc}
K_1^*(x,y) & K_2(x,y)\\
\frac{c}{c^*} K_2^*(x,y) & K_1(x,y)
\end{array}\right)\,.
\end{eqnarray}
Inserting these expressions into \eqref{sol_Psi1} leads to partial differential equations for 
$K_1$ et $K_2$ for which existence and uniqueness of solutions have been established 
in the case of NLSE, see for instance \cite{FT} and references therein.
The important relation to reconstruct $q$ reads
\begin{eqnarray}
cq(x,0)=K_2(x,x)\,.
\end{eqnarray}
Indeed, it can be shown that the kernel element $K_2$ can in principle be obtained from the scattering data via the GLM equation, for $y\le x$,
\begin{eqnarray}
\label{eq_GLM} 
K_2(x,y)=-F(x+y)+\frac{c}{c^*} \iint^x_{-\infty}
ds_1ds_2~F^*(s_1+s_2)K(x,s_1)F(s_2+y)\qquad~~
\end{eqnarray}
where
\begin{eqnarray}
\label{transfo_F}
F(x)=\int_C\frac{dk}{2\pi}\frac{b(k)}{a(k)}
e^{-ikx}\,,
\end{eqnarray}
is a (Fourier) transform of the reflection coefficient $\rho(k)$, ratio of the scattering coefficient $a(k)$ and $b(k)$. 
The contour $C$ goes from $-\infty+i0$ to $\infty+i0$ above all the zeros of $a(k)$ in the upper half-plane. In the case of interest for us, 
we assume that $a(k)$ has no zeros (as we can always remove them with the BTs discussed above) and we can deform the contour down to the real line (from above). 
We iterate the equation to form the Neumann solution for 
$K_2$,
\begin{eqnarray}
K_2(x,x)=-F(2x)-\sum_{n=1}^\infty \left(\frac{c}{c^*}\right)^n\int^x_{-\infty}\dots \int^x_{-\infty} d^{2n}z
F^*(z_1+z_2)\ldots
F^*(z_{2n-1}+z_{2n})\nonumber\\
\times F(x+z_{2n-1})F(z_{2n}+z_{2n-3})\ldots
F(z_4+z_1)F(z_2+x)\,.\qquad
\end{eqnarray}
Using (\ref{transfo_F}), we can perform the integrals over $z_j$ which involve integrals of the form
\begin{eqnarray}
\int^x_{-\infty} dz~e^{i(k_{2j}^*-k_{2j\pm 1})z}=\frac{e^{i(k_{2j}^*-k_{2j\pm 1})x}}{i(k_{2j}^*-k_{2j\pm 1})}\,.
\end{eqnarray}
In the limit where the imaginary part of $k_j$ tends to $0$ from above, this yields the usual prescription 
\begin{eqnarray}
\int^x_{-\infty} dz~e^{i(k_{2j}-k_{2j\pm 1})z}=\frac{e^{i(k_{2j}-k_{2j\pm 1})x}}{i(k_{2j}-k_{2j\pm 1}-i0)}\,.
\end{eqnarray}
Hence we obtain,
\begin{eqnarray}
\label{solu_dev}
q(x,0)=-\frac{1}{c}\sum_{n=0}^\infty
\left(-\frac{c}{c^*}\right)^n\int_{\RR^{2n+1}}\frac{d^{2n+1}k}{(2\pi)^{2n+1}}~
\prod_{j=1}^n\rho^*(-k_{2j})\prod_{j=1}^{n+1}\rho(k_{2j-1})
\frac{e^{-2i\sum\limits_{j=1}^{2n+1}k_jx}}{\prod\limits_{j=1}^{2n} (k_j+k_{j+1}+i0)}\,,\qquad
\end{eqnarray}
We see that this expression is a particular case of the Rosales expansion if we choose the measure $d\lambda(k)$ to be of the form
\be
\label{link}
d\lambda(k)=-\frac{1}{\varepsilon c}\rho(k)\frac{dk}{2\pi}\,,
\ee
where we recall that $\rho(k)=\frac{b(k)}{a(k)}$, with the understanding that it is supported on the real line. Let us just note that 
the factor $2$ in the exponential term comes from the choice of normalisation in $k$ in the auxiliary problem, leading to an extra factor $2$ in the 
corresponding Fourier transform, as already noted in Section \ref{decreasing} when dealing with the quench to the free case.

This observation allows us to take advantage of the result obtained in general by Rosales concerning the resummation of the series representation for the solution. 
The idea is to extend basic vector and matrix representations that would occur in the case of a discrete measure over a finite number of points (the case of a 
pure multisoliton solution) to functions and operator kernels in order to deal with the purely continuous case that concerns us.
Thus, let us define the function
\be
f(k)=e^{-i(kx+k^2t)}\,,
\ee
and the operator $O$ acting on appropriate functions $g(k)$ as
\be
(Og)(k)=\frac{1}{2i\pi}\int_\RR \frac{f^*(k)f(\ell)}{\ell-k+i0}\rho(\ell)g(\ell)\,d\ell\,.
\ee
Note that its conjugate $O^*$ acts as
\be
(O^*g)(k)=\frac{-1}{2i\pi}\int_\RR \frac{f(k)f^*(\ell)}{\ell-k-i0}\rho^*(\ell)g(\ell)\,d\ell\,.
\ee
With these definition, the series representation \eqref{solu_dev} is recognized as a series expansion involving the inverse of the operator $I-\frac{c}{c^*}O^*O$.
More precisely, assuming that this inverse exists, we obtain
\be
\label{operator_form}
q(x,t)=-\frac{1}{c}\int_\RR f(k)\left[(Id-\frac{c}{c^*}O^*O)^{-1}f\right](k)\rho(k)\frac{dk}{2\pi}\,.
\ee
Note that we have incorporated directly the time dependence on the field following from the simple time dependence of the reflection coefficient $\rho(k,t)=e^{ik^2t}\rho(k)$.
The precise analysis of the meaning of the inverse operator is beyond the scope of this paper. The interested reader is referred to \cite{Ros} where some details are given in
the Appendix. The main point is the question of boundedness of the operator $O$ which ultimately relies on the norm properties of the reflection 
coefficient $\rho(k)$. For our purposes, it is sufficient to assume that this expression holds for a given value of $c$ and for a corresponding reflection 
coefficient $\rho(k)$ (pre-quench setup). Indeed, for the quench protocol that we are considering, the important thing to note is that $\rho$ remains unchanged while $c$ is changed 
to $c_0$. If $c_0$ is of the same nature (focusing/defocusing) as $c$ then the $\frac{c_0}{c_0^*}=\frac{c}{c^*}$. Hence, the expression remains well-defined and 
the only change in $q$ is the rescaling that we have already mentioned, as it should. However, if $c_0$ is of a different nature from $c$ then 
$\frac{c_0}{c_0^*}=-\frac{c}{c^*}$. This has two important consequences for the post-quench situation:

$\bullet$ The expression is still well-defined, since what matters is the norm of $\frac{c}{c^*}O^*O$. Very roughly speaking, we are dealing with a series representation of $(1-x)^{-1}$
which we assume to be well-defined, \ie $x$ has norm less than $1$, and we simply require that $(1+x)^{-1}$ be also well-defined and representable by a series.

$\bullet$ The effect on $q$ is no longer a simple rescaling: the change of sign inside the operator is a highly nontrivial transformation on the potential.

The last observation provides us with the missing ingredient to complete our discussion of the map $\cB$ in the general, nontrivial case.

\subsection{Dual quench as a transformation in real space}

Equipped with the two previous notions, here is our strategy to construct the map $\cB$. Assume that the scattering data $S(k,c)$ is 
such that $a(k,c)$ has $N$ simple zeros $k_1,\dots,k_N$. We use the BTs of Section \ref{review_BT} repeatedly to 
remove those zeros while keeping $b(k,c)$ untouched. This produces the purely radiative scattering coefficients
\be
a_r(k,c)=\prod_{j=1}^{N}\frac{k-k_j^*}{k-k_j}a(k,c)\,,~~b_r(k,c)=b(k,c)\,.
\ee
Using these coefficients we can form the reflection coefficient $\rho_r(k,c)=\frac{b_r(k,c)}{a_r(k,c)}$ that enters the Rosales series representation of the 
radiative solution $q_r(x,t)$ as explained in Section \ref{review_Rosales}. 

At this stage, on the one hand, we also know the 
explicit map between $q$ and $q_r$ in terms of the successive Darboux matrices $D_j=k\1-\Sigma_j$ that are used to eliminate the zeros of $a(k)$, cf \eqref{change_potential}
\be
q_r=q+\frac{i}{c}[\sigma_3,\Sigma_1+\dots+\Sigma_N]_{12}\,,
\ee
where the subscript $12$ means that one takes the entry $(12)$ of the $2\times 2$ commutator.

On the other hand, we have the compact formula \eqref{operator_form} for $q_r$ in which we can change the value of the parameter $c$ to $c_0$, 
whether or not $c_0$ is of the same nature as $c$. This gives us $\tilde{q}_r$: the radiative solution with coupling $c_0$ 
that corresponds to the scattering matrix $S_r(k,c)$. There is no easy way to write a transformation formula for this part of the process. 
Formula \eqref{operator_form} resembles an inner product formula so we formally write it as
\be
\label{formal}
q=\langle f|(Id-\frac{c}{c^*}O^*O)^{-1}|f\rangle_c\equiv {\cal O}_c(f)\,,
\ee
where ${\cal O}_c$ is a quadratic form.
With this notation, we can formally write 
\be
\tilde{q}_r={\cal O}_{c_0}\left({\cal O}_{c}^{-1}(q_r)\right)
\ee
Finally, we perform the inverse of the previous BTs to restore the zeros in $S_r(k,c)$ and recover $S(k,c)$. Denoting $\widetilde{D}_j=
k\1-\widetilde{\Sigma}_j$ the successive Darboux matrices used to achieve this, we can write, remembering that we now work with the parameter $c_0$.
\be
\tilde{q}=\tilde{q}_r+\frac{i}{c_0}[\sigma_3,\widetilde{\Sigma}_1+\dots+\widetilde{\Sigma}_N]_{12}
\ee 
Putting everything together, we can write the map $\cB$ of Figure \ref{quench_real1} compactly as
\be
\cB:~~q\mapsto \tilde{q}={\cal O}_{c_0}\left[{\cal O}_c^{-1}\left(q+\frac{i}{c}[\sigma_3,\Sigma_1+\dots+\Sigma_N]_{12}\right)\right]+
\frac{i}{c_0}[\sigma_3,\widetilde{\Sigma}_1+\dots+\widetilde{\Sigma}_N]_{12}
\ee
From our discussion, it is apparent that the crux of the matter when changing the regime of the system (focusing $\leftrightarrow$ defocusing) 
is encapsulated in the object ${\cal O}_{c_0}{\cal O}_c^{-1}$. In the trivial case where the quench preserves the nature of the system, 
${\cal O}_{c_0}{\cal O}_c^{-1}$ becomes $\frac{c}{c_0}I$ and $\widetilde{\Sigma}_j=-\Sigma_j$. We recover consistently 
that $\cB$ is simply the rescaling map $\frac{c}{c_0}I$.

\subsection{Discussion}

From \eqref{intertw}, the dual quench map $\cB$ is related to the original quench map $\cQ_{c,c_0}$ by intertwining with the scattering transform $\cF_{c_0}$. Hence, 
the above suggests that one can use such powerful tools as Darboux-B\"acklund transformations and the Rosales series solution to the problem of quenches in integrable 
classical field theories, based on the example of the NLSE.

In particular, we were led to the construction of the operator $O$ and the related object ${\cal O}_c$ which is 
crucial in capturing the difficult scenario of quenches that do not preserve the nature of the original system. 
This opens the way to treat the case of finite density along the same lines. 
Indeed, in that case, one can also implement the GLM approach as explained in Part One, Chapter II, $\S 7$ of \cite{FT}. The details are a lot more involved 
than in the rapidly decreasing case but the central results are similar: the solution $q(x,t)$ can be reconstructed from the diagonal value of 
a kernel satisfying a GLM equation that one can formally iterate.
Therefore, one could in principle generalize the Rosales type resummation, allowing one to study the effect of a change in the parameter $c$ in the more challenging 
finite density case.

Beyond this extension to the finite density case in the classical case, we now argue that a substantial part of our findings could be quantized and perhaps help 
with the quantum quench problem. Indeed, there is a notion of quantum B\"acklund transformation, at least for integrable systems with a finite number of 
degrees of freedom, which is related to Baxter's $Q$ operator \cite{Bax}, see e.g. \cite{GP,KS,KSS}.
The formal expression \eqref{formal} is also very suggestive of a possible passage to the quantum case. The function $f$ is simply a plane wave so the notation
$|f\rangle$ suggests an asymptotic state. Then, the operator $O$ contains the term $(\ell-k+i0)^{-1}$ which is similar to familiar quantum propagators. 
The whole expression could then represent a resummation of quantum perturbative expansion. This appealing idea is further reinforced by the 
well-known fact that the classical Neumann series solution to GLM goes over to the quantum case as well \cite{TW,CTW,Dav}. The quantized operators corresponding to 
$\rho(k)$ and $\rho^*(k)$ become the generators of the Zamolodchikov-Faddeev algebra \cite{Fad}. The latter algebra was investigated in the context of 
a quantum quench in \cite{SFM}. However, the possible resummation at the quantum level, yielding a quantum analog of ${\cal O}_c$ was not considered. 
We believe that a more detailed investigation of the quantization of our approach could complement the study of \cite{SFM} and 
shed more light on the quantum quench problem. This is beyond the scope of this paper and is left for future investigation.

\section{Conclusion}

In this paper, we have studied the classical quench problem from the viewpoint of the inverse scattering method in the case of the NLSE. 
After reviewing the general statement and introducing the quench map and its relation to the direct scattering transform, we showed explicitly how to solve 
the direct problem on three examples. In the particular case of NLSE, there are two fundamentally distinct regimes: the focusing and defocusing cases. 
It is clear from the examples that quenches can be studied from one regime to another.

Then, we obtained a general representation of the quench map involving 
a sort of conjugation of the scattering data at given coupling $c$, or ``dressing'', by solutions of a higher level scattering problem. The analytic properties 
of the latter were discussed to indicate how they might be used, in conjunction with additional information, to fix the solution uniquely. 

Finally, we explored another route to the problem and considered the dual quench protocol, where the transformation occurs in real space instead of scattering data space. This brings in several natural tools of soliton theory that shed light on the quench problem via the intertwining relation \eqref{intertw}. 
Those tools (GLM equation 
and Rosales solution) are also very likely to prove useful in  establishing a closer relation between our results and existing results for the quantum quench 
problem. Indeed, they are known to survive quantization exactly. In the classical case that occupies us, the crossover through the ``critical'' point $c=0$ 
(the free case) that separates the two regimes is very hard to keep track of in the general case, although we could easily discuss it on the three particular examples.
We showed how the use of B\"acklund transformations and the resummation of the Rosales series expansion provides a means to study this crossover. This was discussed in 
detail for the rapidly decreasing case but the tools involved generalize to the finite density case so that our approach can also be used to cover this case. 

This and other questions related to the present study deserve further investigation to build a more complete theory of classical quenches. This is interesting 
in itself given the close relationship to the inverse scattering method (the direct transform) but also because of the special status of classical integrable systems 
when it comes to quantizing them. In particular, the quantization of certain methods that we have introduced in this paper could prove useful in the 
larger scheme of the study of out-of-equilibrium systems via the toy model scenarios of quenches in integrable systems.

\appendix

\section{Analytic structure of the quench map at a zero of $a(k)$}\label{app}
In section \ref{sectgen}, we derived a set of conditions on the quench map, expressed in Proposition \ref{prop}, or more explicitly in \eqref{fghj1} 
(with solution to the quench given in \eqref{apbp}). The hope is that, seen as analytic conditions in $k$-space, these will fix a one-parameter family 
of solutions (the parameter being $\delta c/c$). The problem of determining this space of solutions is however very nontrivial. As discussed at the end 
of section \ref{sectgen}, a naive counting of the number of degrees of freedom, which excludes an analysis of the singularities around the zeroes of $a$, 
seems to indicate that the space of solution is too vast. However, the structure in $k$-space around the zeroes of $a$ is more rigid. The purpose of this 
appendix is to study, very briefly, the form of some additional constraints that arise.

Assume that $a(k)$ has a zero of order $j>0$ at $k=k_0$. Since $a=\det(\Phi_>)$, this implies that the product of the eigenvalues of $\Phi_>$ has such a zero. 
It is natural to assume that one of the two eigenvalues display this pole, that is, the other eigenvalue stays nonzero. Let us keep the $x$ dependence implicit. 
As mentioned in section \ref{sectgen}, at $k=k_0$ we have 
\be
\Phi_>(k_0)B=0
\ee
where
$B = (-b_0,1)^T$ for some $b_0\in \CC$. In a neighbourhood of $k_0$, $\Phi_>(k)$ is analytic so we can write, in terms of 
$\delta: = k-k_0$ 
\be
\Phi_>=\sum_{k=0}^\infty A_k\delta^k
\ee
with 
$A_kB=0$, $k=0,\dots,j-1$ and $A_0\neq 0$ in view of our assumption that only one of the two eigenvalues of $\Phi_>$ has a zero at $k_0$.
We can also write
\beq
	\frc1 a = \delta^{-j}\alpha(\delta)
\eeq
where $\alpha(\delta)$ has a Taylor series expansion in $\delta$ with $\alpha(0)\neq0$.

The claim we make is that the post-quench quantity $\Theta_{\ep>}P_{-\ep}$ has an analytic expansion around $k_0$ that is constrained by the pre-quench data $\Phi_>$
 and $\Phi_<^\dagger$ only, for any $\ep$ and for any fixed value of the position $x$ 
(say $x=0$). Specifically, we now assume that the function $\Theta_{\ep>}$ has a pole of order $\ell\geq0$ at $k=k_0$, and an expansion with 
coefficients that are differentiable in 
$x$. With the linearly independent (in fact orthogonal) vector $\t B = (1,b_0^\star)^T$, and based on the arguments presented in section 
\ref{sectgen} showing that it must be proportional to $\delta^{-\ell} (B + O(\delta))$\footnote{Note that $\Theta_{\ep>}\PP_{-\ep}$ 
can be identified with $m(x)$ in proposition \ref{prop}.}, we may write\footnote{We use a 
slight abuse of language, as $\Theta_{\ep>}P_{-\ep}$ is a $2\times 2$ matrix;  we mean the nonzero vector of this matrix.}
\beq\label{thetaTaylor}
	\Theta_{\ep>}\PP_{-\ep} = \delta^{-\ell}
	(\theta(\delta)B + \delta\t\theta(\delta)\t B)
\eeq
where $\theta(\delta)$ and $\t\theta(\delta)$ both have Taylor series expansions in $\delta$, with $\theta(0)\neq0$. Then, the statement is that
 $\t\theta$ is completely determined 
as a power series in $\delta$ in terms of $\theta$, $\Phi_>$, $\Phi_<^\dagger$, $q(x)$ and the relative quench ratio $C = \frc{\delta c}{c}$. 
To see this, note that the differential form of \eqref{tek},
\[
	\p_x \Theta_{\ep>}\PP_{-\ep} = \frc{C}a U_>\Theta_{\ep>}\PP_{-\ep}\,,
\]
implies
\be
\partial_x\theta\,B+\delta\partial_x\t\theta\,\t B=C\alpha \delta^{-j}U_>(\theta\,B+\delta\t\theta\,\t B)
\ee
and hence,
\be
\partial_x\theta\,B^\dagger\,B=C\alpha \delta^{-j}\,(\theta\,B^\dagger U_>B+\delta\t\theta\,B^\dagger U_>\t B)
\ee
which yields
\be
\label{eq_theta}
\delta \,B^\dagger U_>\t B\,\t\theta=\frac{\delta^{j}}{C\alpha}\partial_x\theta\,B^\dagger\,B-\theta\,B^\dagger U_>B\,.
\ee
The claim now follows from two facts
\begin{itemize}
\item The right-hand-side is of order (at least) $\delta$;

\item $B^\dagger U_>\t B|_{\delta=0}\neq 0$, which ensures that $(B^\dagger U_>\t B)^{-1}$ is analytic around $\delta = 0$.

\end{itemize}
The first term in the right-hand side of \eqref{eq_theta} is clearly of order $\delta$ (recall that $\alpha(0)\neq 0$ and $j\ge 1$). In the second term, $B^\dagger U_>B$ equals
\be
B^\dagger \Phi_<^\dagger\, U\, \sum_{k\ge j} A_kB\,\delta^k\,,
\ee 
which is also at least of order $\delta$ given that $\Phi_<^\dagger$ is analytic around $\delta=0$ as well. That proves the first fact. Now, if 
\be
B^\dagger U_>\t B|_{\delta=0}=0
\ee
then
\be
\text{either}~~\Phi_>(0)\t B=0~~\text{or}~~B^\dagger \Phi_<^\dagger(0)=0\,,
\ee
given that $\det U=|q(x)|^2\neq 0$. The first possibility means that $A_0 \t B=0$, which is a contradiction with $A_0 B=0$ and $A_0\neq 0$. 
The second possibility also leads to a contradiction for the same reason: we have $\t B^\dagger \Phi_<^\dagger(0)=0$ but it is natural to assume 
$ \Phi_<^\dagger(0)\neq 0$ in a generic nontrivial case. That concludes the argument which shows that the final expression for $\t\theta$  
\be\label{ttheta}
\t\theta=\frac{\delta^{j-1}}{C\alpha}\,\frac{B^\dagger\,B}{B^\dagger U_>\t B}\partial_x\theta-\frac{B^\dagger U_>B}{\delta\,B^\dagger U_>\t B}\theta\,.
\ee
is a power series in $\delta$ involving only the known quantities mentioned above.

\end{document}